\documentclass[twocolumn,tradiabstract]{aa} 

\usepackage{natbib}
\usepackage{graphicx}
\usepackage{amsmath}

\usepackage{xcolor}
\newcommand{\pd}[2]{\frac{\partial #1}{\partial #2}}

\newcommand\mancha{\textsc{Mancha3D~}}

\begin{document}

\title{Three-dimensional simulations of solar magneto-convection including effects of partial ionization}
\titlerunning{Magneto-convection including partial ionization}

\author{E. Khomenko\inst{1,2}, N. Vitas\inst{1,2}, M. Collados\inst{1,2}, A. de Vicente\inst{1,2}}
\authorrunning{E. Khomenko et al.}

\institute{Instituto de Astrof\'{\i}sica de Canarias, 38205 La Laguna, Tenerife, Spain
\and Departamento de Astrof\'{\i}sica, Universidad de La Laguna, 38205, La Laguna, Tenerife, Spain}

\date{Received; Accepted }

\abstract {Over the last decades, realistic 3D radiative-MHD simulations have become the dominant theoretical tool for understanding the complex interactions between the plasma and the magnetic field on the Sun. Most of such simulations are based on approximations of magnetohydrodynamics, without directly considering the consequences of the very low degree of ionization of the solar plasma in the photosphere and bottom chromosphere. The presence of large amount of neutrals leads to a partial decoupling of the plasma and the magnetic field. As a consequence of that, a series of non-ideal effects (ambipolar diffusion, Hall effect and battery effect) arises. The ambipolar effect is the dominant one in the solar chromosphere. Here we report on the first three-dimensional realistic simulations of magneto-convection including ambipolar diffusion and battery effects. The simulations are done using the newly developed \mancha code. Our results reveal that ambipolar diffusion causes measurable effects on the amplitudes of waves excited by convection in the simulations, on the absorption of Poynting flux and heating and on the formation of chromospheric structures. We provide a low limit on the chromospheric temperature increase due to the ambipolar effect using the simulations with battery-excited dynamo fields.}


\maketitle

\section{Introduction} %

The solar chromosphere is a boundary layer between the interior and the exterior of the Sun. The plasma in the solar photosphere and chromosphere is only partially ionized. In the recent few years it has been repeatedly demonstrated that processes related to the non-ideal plasma behavior due to the presence of the neutral plasma component may be the key ones to solve the problem of chromospheric heating, dynamics and fine structure \citep{DePontieu1998, Goodman2000, Judge2008,  Krasnoselskikh2010, Goodman2010, Goodman2011, Song+Vasyliunas2011, MartinezSykora+etal2012, Khomenko+Collados2012, Shelyag+etal2016}. 

The chromosphere hosts shock waves and current layers. Magnetic field lines, rooted in the photosphere, are constantly perturbed by convective plasma motions, which serve as an energy reservoir. A significant part of energy of these motions is transported to the chromosphere in the form of different types of waves. However, not all wave types are able to reach and deposit their energy at the chromosphere. According to the mode conversion theory \citep{Cally2006, Cally+Goossens2008}, acoustic $p$-modes propagating in subsurface layers can be transformed into fast magneto-acoustic and Alfv\'en waves when reaching the magnetically dominated layers higher up. Alfv\'en waves created by this mechanism might be able to reach the chromosphere more easily. This mode transformation mechanism is confirmed by numerical simulations \citep{Khomenko+Collados2006, Khomenko+Cally2012, Felipe2012} and is recognized to be responsible for the formation of high-frequency acoustic halos surrounding active regions \citep{Khomenko+Collados2012, Rijs2015, Rijs2016}. 

Magnetic waves reaching the chromosphere may create perturbations in magnetic field, i.e. produce perturbations of currents. These currents can be dissipated via either classical Ohmic diffusion or ambipolar diffusion, related to the presence of neutrals. At chromospheric layers, as a consequence of incomplete collisional coupling, ions would move under the action of the magnetic Lorentz force and would be constantly disrupted by neutrals that do not sense this force directly, but only through collisions with ions. The magnetic field frozen in the charged plasma component would diffuse due to this differential motion between ions and neutrals. This friction creates dissipation and therefore allows to transform magnetic energy into heat. Mathematically, when the single-fluid approximation is adopted, the above effect results in large ambipolar diffusion coefficients \citep{Khomenko+Collados2012, Khomenko+etal2014b} and the internal energy of the plasma would increase proportional to $de_{\rm int}/dt=\eta_A J_\perp^2$. The ambipolar diffusion effect removes currents perpendicular to the magnetic field, approaching a force-free field in the relaxed state \citep{Leake+Arber2006, Arber2007}.

The dissipation of static currents at the borders of chromospheric flux tubes has been shown to be efficient to balance the radiative losses of the chromosphere \citep{Khomenko+Collados2012, Khomenko+Collados2012b}. However, currents created dynamically by waves and motions are found to be even more effective. There are theoretical works and idealized numerical simulations suggesting that dissipation of Alfv\'en waves due to effects of ion-neutral interactions is an important source of chromospheric heating \citep[][among the recent ones]{Goodman2000, Goodman2011, Shelyag+etal2016}. In particular, \citet{Goodman2000} concluded that the chromosphere is heated by flux tubes with photospheric magnetic field strength between 700 and 1500 G. \citet{Shelyag+etal2016} considered a more complete model where acoustic waves were generated outside of a magnetic flux tube, then entered the flux tube, and double converted into the fast magnetic waves and then into Alfv\'en waves. In their example up to 80\% of the magnetic Poynting flux of waves could be absorbed and converted to heat. In a continuation of this work \citet{Przybylski+etal2017} demonstrated that this absorption depends non-trivially on the frequency and is a highly non-linear process in which the amount of absorption depends on the wave amplitude. Other studies concluded that compressible dissipation in shocks is as efficient (or more) as dissipation due to the ambipolar diffusion mechanism \citep{Goodman2010, Arber2016}.

Realistic numerical simulations are a necessary next step to provide quantitative conclusions about the importance of neutrals in the chromospheric energy balance, and to evaluate the role of other mechanisms mentioned above, such as shocks.  Realistic 2D  numerical simulations including ambipolar diffusion have been employed by \citet{MartinezSykora+etal2012, MartinezSykora2016, MartinezSykora2017}. \citet{MartinezSykora+etal2012} showed that the temperature in cool chromospheric bubbles produced during the nearly-adiabatic expansion of the material reaching upper layers does not drop to such low values as in the models without ambipolar diffusion. \citet{MartinezSykora2016, MartinezSykora2017} claimed that spicules can be produced due to the effects of ambipolar diffusion by allowing the field lines to reach higher, and that chromospheric fibrils and the magnetic field can become visibly misaligned due to the same effect. Apart from wave fronts, vortex motions have been recently identified as another way of energy transport to the upper layers \citep{Moll2011, Moll2012, Wedemeyer2012, Shelyag2013, Kitiashvili2013, Wedemeyer2014}. When three-dimensional simulations of magneto-convection are employed, it has been shown that vortices are formed differently and have different properties in non-magnetized and magnetized areas. Those in magnetized areas reach higher and can act as channels energetically connecting the photosphere and the chromosphere. 

The aim of the present work is to advance the investigation of the ion-neutral effects on the energy balance of the solar chromosphere. To that aim, we have performed 3D simulations of dynamo and magneto-convection including the effect of ambipolar diffusion (as a main non-ideal effect), using the generalized Ohm's law. We statistically compare pairs of simulation runs with/without ambipolar diffusion. Results regarding the Poynting flux absorption as a function of frequency and magnetic structure are presented. We also study the formation of fine structures. The effects on the average chromospheric temperature increase are discussed.

\begin{figure*}[t]
\begin{center}
\includegraphics[width = 8cm]{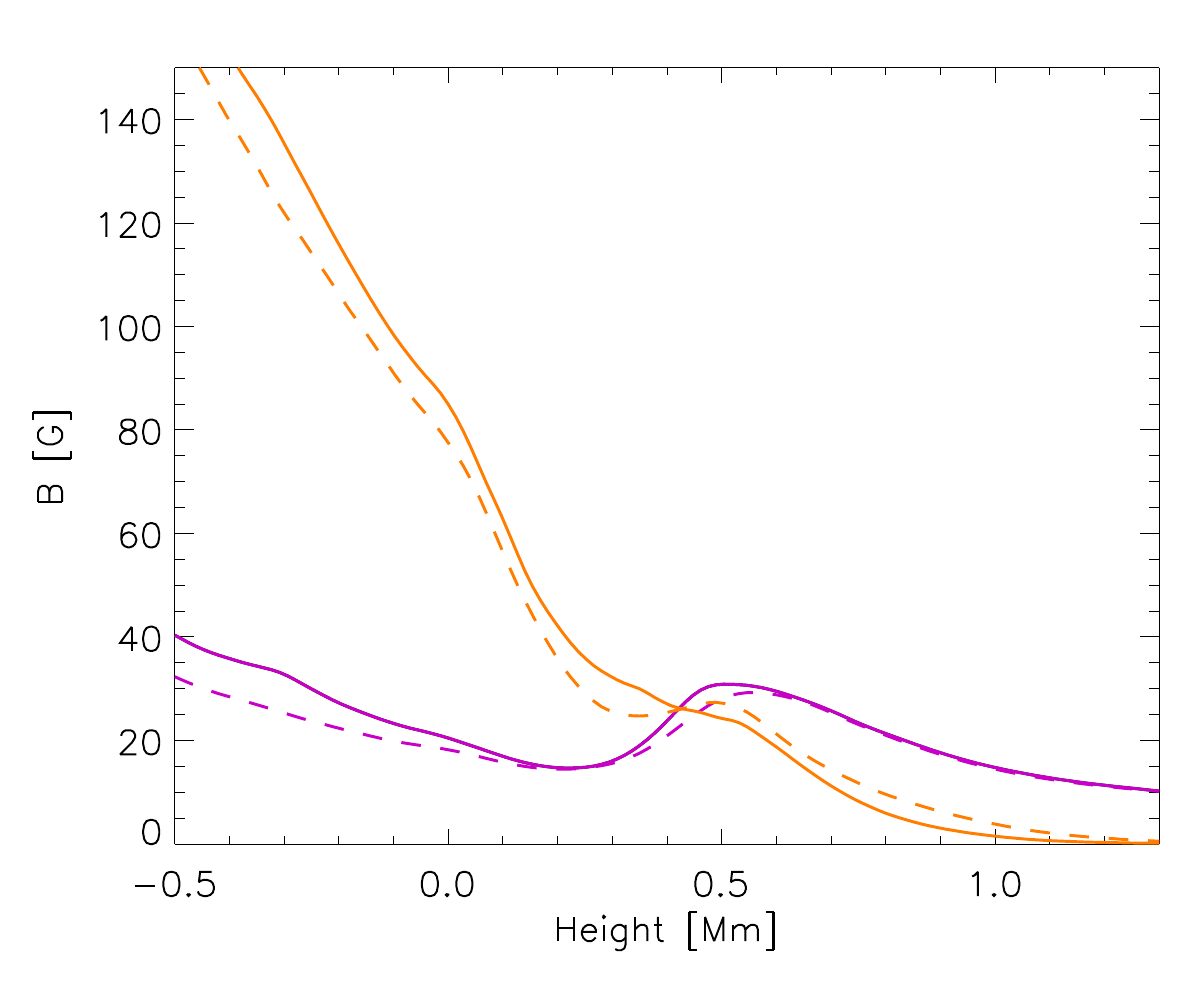}
\includegraphics[width = 8cm]{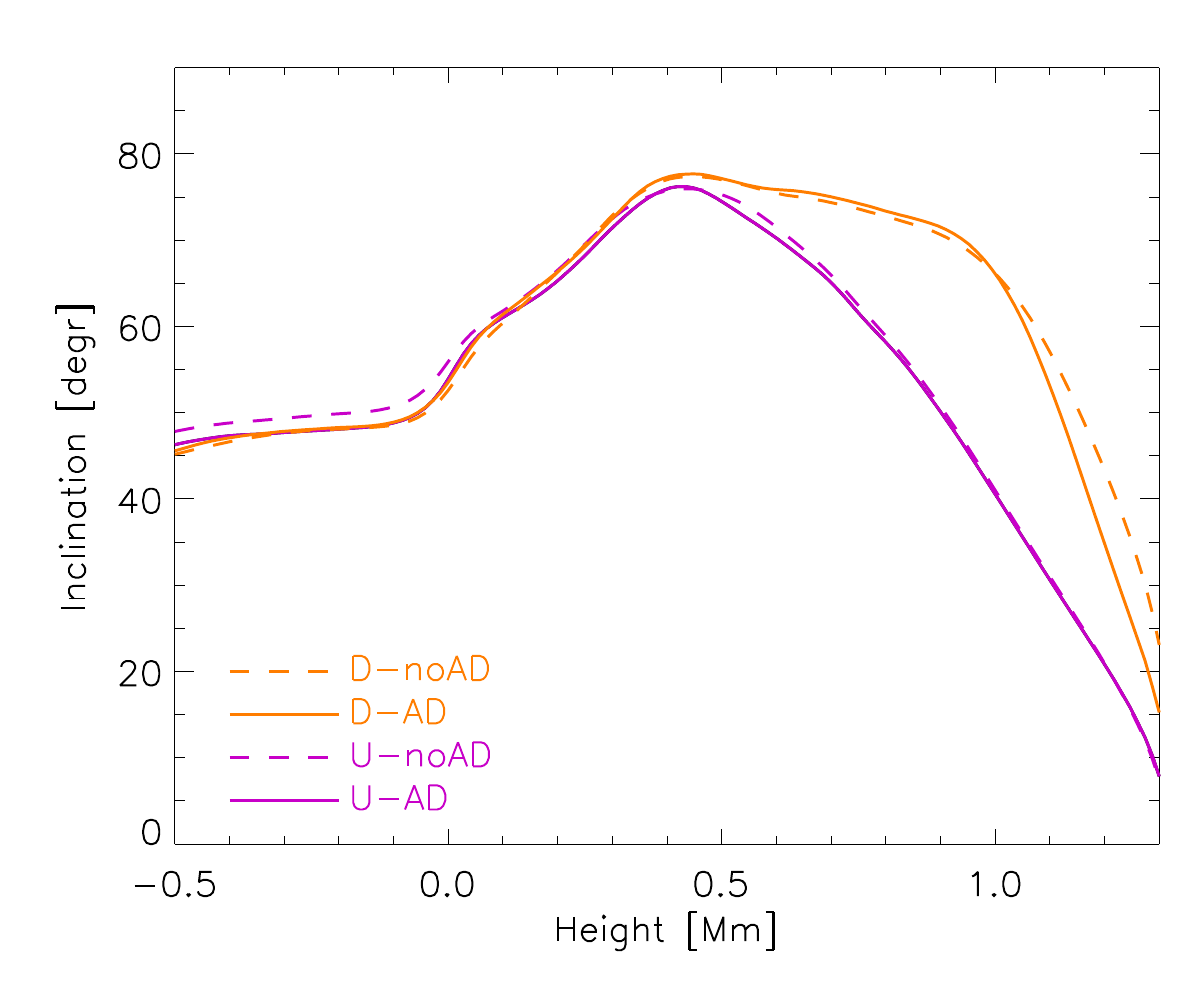}
\end{center}
\caption{Left panel: time- and space-averaged modulus of the magnetic field strength as a function of height in the four simulation runs. Right panel: time- and space-averaged magnetic field inclinations with respect to the vertical direction. Orange solid: D-AD; orange dashed: D-noAD; violet solid: U-AD; violet dashed: U-noAD.  The time averaging interval is 50 min.}
\label{fig:bmean}
\end{figure*}

\section{Magneto-convection simulations by \mancha}

The simulations presented in this paper were done with the \mancha code \citep{Khomenko+Collados2006, Felipe+etal2010, Gonzalez-Morales+etal2017}. \mancha solves the non-ideal non-linear equations of single-fluid magnetohydrodynamics, with equilibrium being explicitly removed from the equations. The code treats the most important effects derived from the presence of neutral atoms in the solar plasma: ambipolar diffusion, Hall effect and battery effect \citep{Khomenko+Collados2012b}, assuming that plasma is strongly colisionally-coupled. In this approximation, well justified in the photosphere and low chromosphere, the non-ideal effects manifest themselves through additional terms in the generalized induction equation, with the corresponding repercussion on the energy conservation equation (Eqs. \ref{eq:induction} and \ref{eq:energy} below), see \citet{Khomenko+etal2014, Ballester2017}. The numerical treatment of these terms is described in \citet{Gonzalez-Morales+etal2017}. 

In this work we neglect the Hall effect and therefore the only two non-ideal effects treated in this study are the ambipolar diffusion and the battery effect. The following equations are solved in the code:

\begin{equation}
\label{eq:continuity} 
\pd{\rho_1}{t} + \mathbf{\nabla} \cdot \left( \rho\mathbf{v} \right) =  \left( \pd{\rho_1}{t}\right)_{\rm diff},
\end{equation}

\begin{eqnarray} \label{eq:momentum} 
\pd{\rho\mathbf{v}}{t} &+& \mathbf{\nabla}\cdot \left[\rho\mathbf{v} \mathbf{v} + \left(p_1+ \frac{\mathbf{B}_1^2 + 2\mathbf{B}_1 \cdot \mathbf{B}_0}{2 \mu_0} \right) \mathbf{I} \right] + \nonumber \\
&+&\mathbf{\nabla}\cdot \left[\frac{1}{\mu_0} \left(\mathbf{B}_0 \mathbf{B}_1-\mathbf{B}_1 \mathbf{B}_0-\mathbf{B}_1 \mathbf{B}_1 \right) \right] + \nonumber \\
&=& \rho_1 \mathbf{g} + \left( \pd{\rho\mathbf{v}}{t}\right)_{\rm diff} ,
\end{eqnarray}

\begin{eqnarray} \label{eq:induction}
\pd{\mathbf{B}_1}{t}  &=&  \mathbf{\nabla}\times \left[\mathbf{v}\times \mathbf{B} +\frac{\mathbf{\nabla}p_e}{e n_e} + \frac{\left(\eta_A\mathbf{J} \times \mathbf{B} \right) \times \mathbf{B} }{|B|^2}  \right] + \nonumber \\
&+& \left(\pd{B_1}{t}\right)_{\rm diff}, 
\end{eqnarray}

\begin{eqnarray} \label{eq:energy}
\pd{e_1}{t} &+& \nabla \cdot \left[ \mathbf{v}\left(e + p + \frac{|\mathbf{B}|^2}{2 \mu_0} \right) - \frac{\mathbf{B}(\mathbf{v} \cdot \mathbf{B}) }{\mu_0} \right]   =  \nonumber \\
&=& \rho \left(\mathbf{g} \cdot \mathbf{v}\right)  +\mathbf{\nabla}\cdot \left[ \frac{\mathbf{B} \times (\eta_A \mathbf{J_\perp}) }{\mu_0}   + \frac{\mathbf{\nabla}p_e \times \mathbf{B}} {e n_e \mu_0}    \right] \nonumber \\
&+& Q_{\rm R} + \left( \pd{e_1}{t}\right)_{\rm diff}  
\end{eqnarray}
where generally, $\rho=\rho_0+\rho_1$, $p=p_0+p_1$, $\mathbf{B}=\mathbf{B}_0+\mathbf{B}_1$, $\mathbf{v}=\mathbf{v}_1$, and  $e=e_0+e_1=(\rho_0+\rho_1)\mathbf{v}^2/2 + (\mathbf{B}_0+\mathbf{B}_1)^2/2/\mu_0 + (e_{\rm int 0}+e_{\rm int 1})$ is the total energy. In the code we have a freedom to set variables as background (index ``0'') or perturbation (index ``1''), with the only requirement that the background variables fulfill the condition of (M)HS equilibrium. In the experiments performed here the initial equilibrium was chosen to be purely hydrostatic, so that $B_0=0$ and all the magnetic field vector is contained in the variable $B_1$. The terms subscripted with ``diff'' are numerical diffusion terms required for numerical stability. These terms are computed according to \citet{Vogler2005, Felipe+etal2010}, see also the detailed description in the PhD Thesis of A. V\"ogler\footnote{https://www.mps.mpg.de/phd/theses/three-dimensional-simulations-of-magneto-convection-in-the-solar-photosphere.pdf}.

The ambipolar diffusion coefficient is calculated in units of $[ml^3/tq^2]$ according to the following expression:
\begin{eqnarray} \label{eq:etaa}
\eta_A =\frac{\xi_n^2 |B|^2}{\alpha_n}.
\end{eqnarray}
with the collisional parameter $\alpha_n$ defined as:
\begin{eqnarray}
\alpha_n &=& \sum_b\rho_e\nu_{e; bn} + \sum_{a,i=1}\sum_b\rho_{ai}\nu_{ai; bn}
\end{eqnarray}
with the summation going over all available chemical species of neutrals and singly ionized ions. The collisional frequencies $\nu_{e; bn}$, $\nu_{ai; bn}$ are defined in \citet{Braginskii1965, Draine1986, Lifschitz, Rozhansky}, see also the review by \citet{Ballester2017}.

The system of equations is closed with an equation of state (EOS) for the solar chemical mixture given by \cite{1989GeCoA..53..197A}. The internal energy ($e_{\rm int}$), density ($\rho$) and electron pressure ($p_e$) are precomputed for temperature ($T$) - pressure ($p$) grid following the algorithm of \cite{1967MComP...7....1M} based on the Saha equation. The EOS takes into account the effects of the first and second atomic ionization for all included elements ($Z \le 92$), and formation of hydrogen molecules. The ionization fraction is obtained self-consistently with the rest of thermodynamical variables providing all necessary variables to compute the neutral fraction, $\xi_n=\rho_n/\rho$ and the ambipolar diffusion coefficient via Eq. \ref{eq:etaa}. The EOS results are stored in lookup tables giving efficient conversion method between thermodynamic quantities and a possibility to use different EOS realization (with different chemical composition or including non-ideal terms like electron degeneracy and pressure ionization) that are relevant for other simulations stretching deeper into the convection zone.

The radiative loss term $Q_{\rm R}$ is computed by solving the non-gray Radiative Transfer (RT) equation, assuming Local Thermodynamic Equilibrium (LTE), using domain decomposition, as in \citet{Vogler2005}. The wavelength dependence of the emission and absorption coefficients is discretized using the opacity binning method \citep{Nordlund1982}.  In the runs considered in this paper we used one single bin, i.e. ``grey'' approximation. The angle integration is done using quadrature with 3 rays per octant. The formal solver used in the RT module is based on the short-characteristics method by \cite{Olson+Kunasz1987}. The LTE approximation limits the accuracy of the RT module in simulations extending above the photosphere. In the optically thin corona, the code uses Newtonian cooling approximation.  Nevertheless, keeping these limitations in mind, here we have computed models extending somewhat higher to the lower chromosphere, reaching approximately 1400 km above the solar surface, as detailed in the section below.

\begin{figure*}[t]
\begin{center}
\includegraphics[width = 18cm]{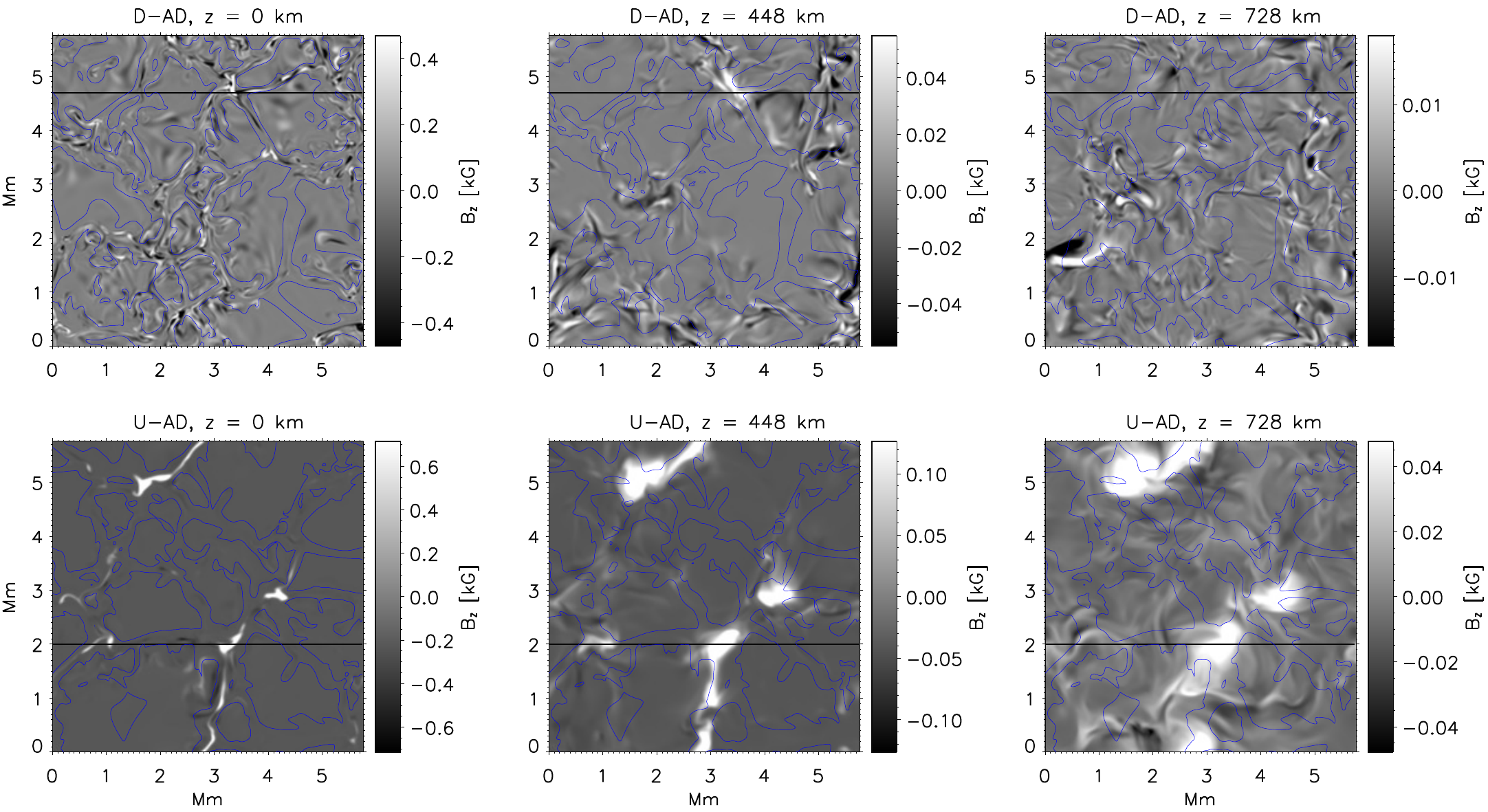}
\end{center}
\caption{Vertical magnetic field at three selected heights for a snapshot of the D-AD run (upper row) and U-AD run (bottom row). The heights are indicated at the title of each panel. Notice the complex structure of fields in the D run, and the expansion of the magnetic field structures with height in both runs. Contours indicate the locations with zero vertical velocity at $z=0$, highlighting the structure of granulation. The horizontal black line marks the location of cuts shown in Figure \ref{fig:tcut} below.}
\label{fig:bexample}
\end{figure*}

\begin{figure*}
\begin{center}
\includegraphics[width = 14cm]{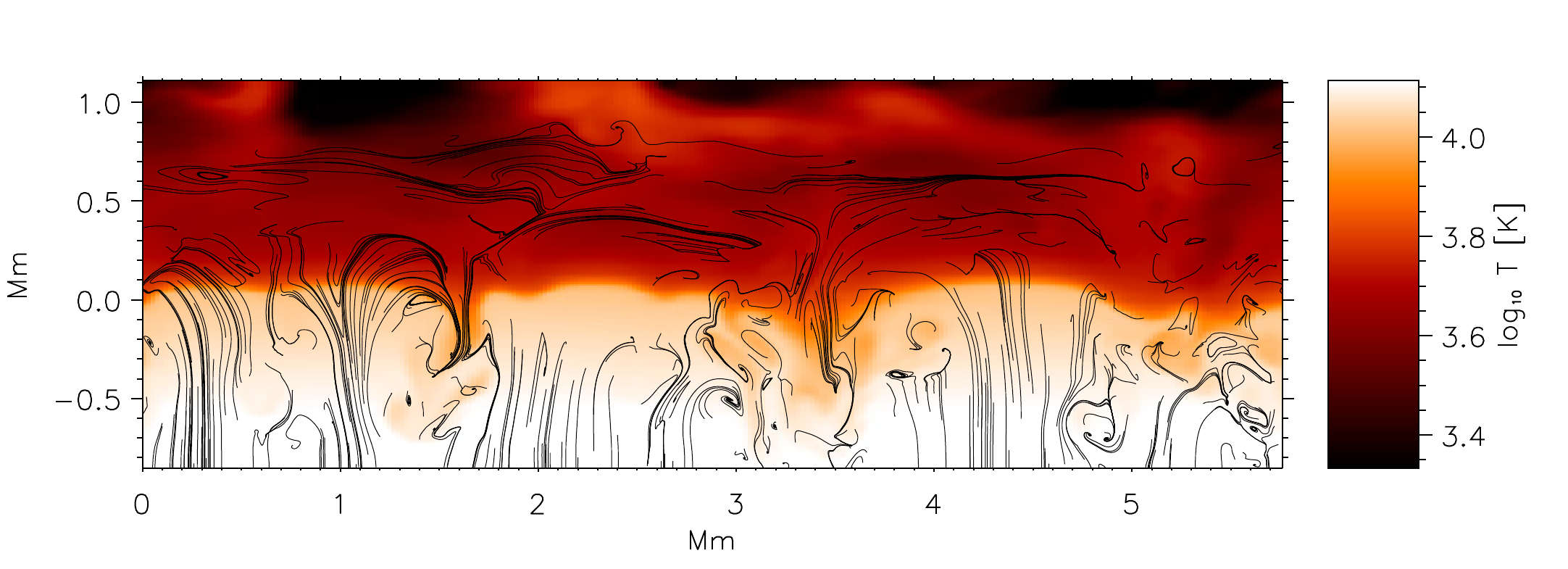}
\includegraphics[width = 14cm]{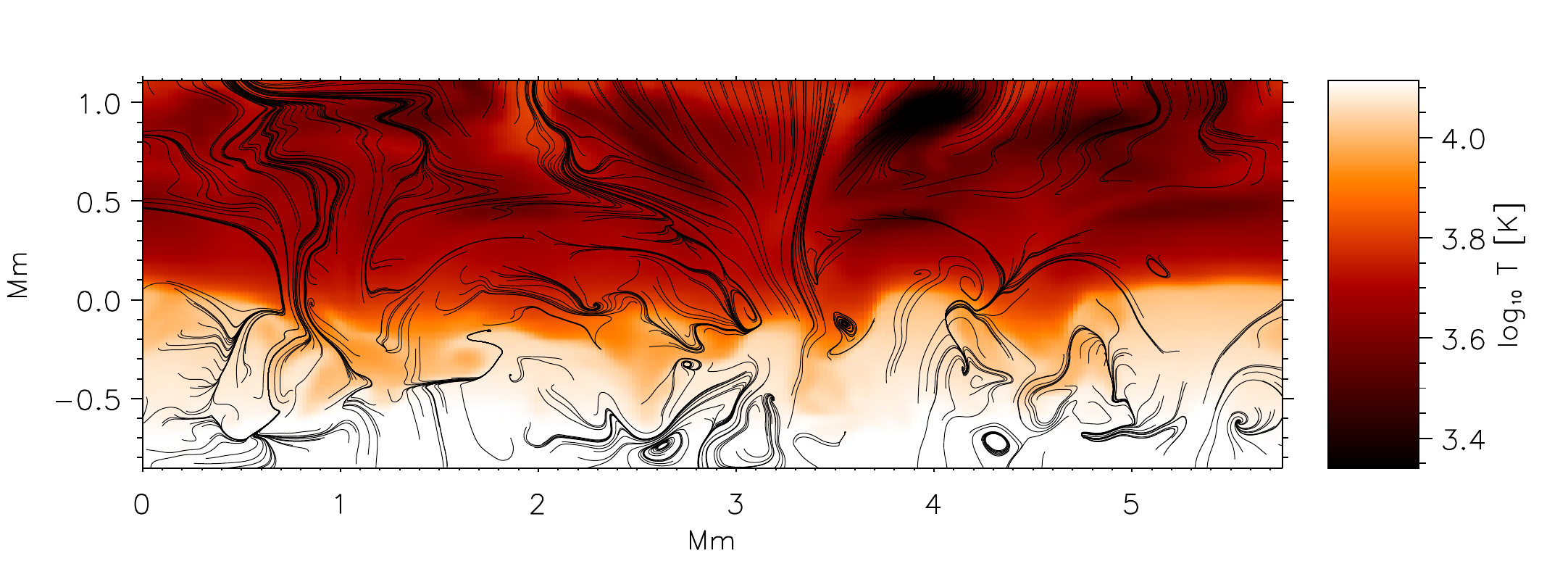}
\end{center}
\caption{Cuts through the simulation snapshots over horizontal $x$ and vertical $z$ directions. The locations of the cuts are marked in Figure \ref{fig:bexample}. The color image shows the temperature. The black lines are the projections of the magnetic field lines into the $x-z$ plane. Only the field lines with the average field strength above 5 G are drawn.Top panel: D-AD run; bottom panel: U-AD run.}
\label{fig:tcut}
\end{figure*}

\begin{center}
\begin{table}
\begin{center}
\caption{Parameters of the simulation runs} 
\begin{tabular}{ccc}
\hline ID &  Ingredients & Time with AD \\ \hline
D-AD & battery, dynamo, AD & 150 min \\
D-noAD & battery, dynamo & $-$ \\
\hline
U-AD & unipolar $B=10$ G, AD & 206 min  \\
U-noAD & unipolar $B=10$ G & $-$ \\
\hline
\end{tabular}
\end{center}
\end{table}
\end{center}

\subsection{Details of the runs}

Here we discuss four runs of magneto-convection simulations. To initiate the convection we started with a plane-parallel atmosphere that combines a model of solar convection zone by \citet{Spruit1974} below the surface with the HSRA model \citep{hsra} above the surface. The two models are joint in a smooth way imposing the hydrostatic equilibrium and consistency with the equation of state used in the code.
The box size is 5.8$\times$5.8$\times$2.35 Mm$^3$ covered by 288$\times$288$\times$168 uniformly distributed grid points, i.e. with a horizontal sampling of 20 km and a vertical sampling of 14 km. The height range includes approximately 0.95 Mm below the surface and about 1.4 Mm above the surface. ``Random'' noise was introduced to the internal energy to initiate the instability. The horizontal boundaries are periodic and the upper boundary is closed for mass flows. We set symmetric boundary conditions  (zero gradient) in internal energy and density variables at the top boundary, and the temperature is computed through the equation of state. We do not set any specific condition to keep the temperature at a given value at the top boundary, allowing shock viscous dissipation, magnetic field related mechanisms, and radiative losses to set the structure of the atmosphere. The bottom boundary is open for the mass flow with automated control of the fluctuation of the total mass and the total radiative output \citep[similar to][]{Vogler2005}, and with zero magnetic field inflow. This way the simulation maintains a required value of outgoing flux and keeps approximately constant the mass of the domain.

The initial equilibrium atmosphere is set in a way that the outgoing flux is equal to the mean solar value of 6.3$\times10^{7}$ J m$^{-2}$s$^{-1}$. Since convection is not initially developed, it does not provide convective energy transport and there is only energy transport by radiation. Therefore, at the beginning of the simulation, the outgoing flux starts to decrease reaching a minimum of about 2.3$\times10^{7}$ J m$^{-2}$s$^{-1}$ after about 9 minutes of solar time. The mean temperature in the simulation box, given by $T=T_0+T_1$ drops down in this initial phase.  At this stage, convection is developed enough and becomes efficient channel for energy transport. Also, the downflowing material has already reached the lower boundary. Altogether, convection leads to a smooth recovery of the value of the outgoing flux back to the solar value, and also results in a temperature increase. In total, this initial phase takes about 20 min of solar time.  We then run the purely hydrodynamical (HD) convection for another 3.4 hours of solar time to make sure it reached a stationary regime. 

At this point we take a snapshot of stationary HD convection and perform one of the following actions: (1) switch on the battery term in Eqs. \ref{eq:induction} and \ref{eq:energy}; (2) introduce a constant vertical unipolar field of 10 G strength into the simulation box. We will refer to the first case as ``dynamo'' (D) run and to the second case as ``unipolar'' (U) run.

The battery term in the D run generates the small-scale dynamo seeds with a strength of $10^{-6}$ G, see \citet{Khomenko+etal2017}. Its continuous action, together with dynamo amplification, provides the magnetization of the model with  $\sim10^2$ G mean field strength at the $\tau_5$=1 surface, after reaching the saturated regime in about 2 hours of solar time \citep[see Figure 3 in][]{Khomenko+etal2017}. We run D simulation for 4.2 hours of solar time in total after switching on the battery term, with about 2 hours in the stationary saturated dynamo regime. 

In the U run, the field is quickly moved into intergranular lanes through flux expulsion, where it forms kG concentrations, as seen many times in the simulations reported in the literature \citep[see e.g.,][]{Vogler2005}. We let the U simulation run for 82 min of solar time to reach a new stationary state. 

Finally, we take a snapshot of either (1) stationary ``dynamo'' (D) run or (2) stationary ``unipolar'' (U) magneto-convection run, and switch on the ambipolar term (referred as AD later) in equations \ref{eq:induction} and \ref{eq:energy}. We then let the D simulation run for another 150 min of solar time with the ambipolar diffusion term on. During the last 50 min we saved snapshots every 20 sec and used them for the analysis in the current paper. We label this simulation as D-AD, see Table 1. Similarly, we let the U simulation run 206 min with AD on, and saved snapshots every 20 sec during the last 50 min. This simulation is labeled as U-AD, see Table 1.

In parallel, we kept running D and U simulations with exactly the same setup but with ambipolar diffusion switched off (D-noAD and U-noAD from Table 1). These runs are used for comparison purposes. Since our radiative transfer losses calculation for the chromosphere include several approximations (as LTE), we cannot make conclusions about the average temperature structure in the chromosphere from such runs. We rather do a statistical comparison between the temperature and other parameters in the runs with and without ambipolar diffusion.

The \mancha code uses temperature fix based on internal energy in order to prevent the temperature in the domain go below a certain limit (about 2000 K). The temperature in the D-AD run needs to be fixed in 0.0045\% of points, while in the D-noAD run it is in 0.0057\% of points. In the U runs the percentage of points to fix is similar in the AD and noAD case, 0.0004\%.

The resulting distributions of the time-averaged magnetic field strength and inclination with height in all four runs are shown in Figure \ref{fig:bmean}. In the D runs (orange solid and dashed lines), the field smoothly decreases its strength with height from hG values below the photosphere to a few G in the chromosphere, with a secondary maximum around 0.5 Mm in the photosphere. This maximum is produced by strongly inclined fields, connecting complex bi-polar structures produced in these simulations, see Figure \ref{fig:tcut} and the corresponding discussion later on. The average inclination of the field in the D-runs, shown at the right panel of Fig. \ref{fig:bmean} (orange curve),  reflects this behavior. The field is on average more vertical at heights below the photosphere (average inclinations around 40-50 degrees). Between about 0.2 and 1 Mm, the average inclination is 70--80 degrees. The field strength becomes extremely weak in the D-runs above 1 Mm, and its inclination tends to the vertical again, imposed by the boundary condition in our simulations. In the U run, a similar tendency is observed with the field decreasing with height and with a more pronounced secondary maximum (violet solid and dashed lines in Fig. \ref{fig:bmean}). The mean field in the U simulations is lower than in the D simulations in the sub-surface layers and in the photosphere, but is larger in the chromosphere. This happens because more chaotic and complex fields produced in the dynamo simulations are connected and form loops at lower heights, compared to the relatively unipolar U runs. The field expands with height and becomes more horizontal in the U-runs around 0.4--0.5 Mm (inclinations close to 80 degrees in Fig. \ref{fig:bmean}, right panel), but then it quickly becomes vertical again. The comparison between AD and noAD runs reveals that the mean field strength is slightly lower in noAD runs in the sub-surface layers and in the low photosphere. In the chromosphere, the mean $B$ in noAD run is higher than in the AD one for the ``dynamo'' simulations, with no appreciable difference for the ``unipolar'' simulations.  
 
The horizontal distribution of the vertical magnetic field component is shown in Figure \ref{fig:bexample} for the D-AD and U-AD runs. As expected, the field has an extremely complex structuring in the D-AD run with an almost balanced mixture of positive and negative polarities. The magnetic structures are mostly rooted in intergranular lanes at photospheric level ($z=0$ km) and are expanding with height. This expansion is more evident in the U-runs. In the D-runs, one can observe the expansion comparing the field maps at 0 and 488 height (left and middle image in Fig. \ref{fig:bexample}, top).  At the lower chromosphere ($z=$728 km) a layer of strongly inclined canopies is produced in both simulations. These canopies are evident from the rightmost panels of Fig. \ref{fig:bexample} by the presence of the fields filling a large fraction of space, and not only intergranular zones, as in the photosphere. The differences in the overall structure of the field in both types of simulations are better visualized in Figure \ref{fig:tcut}, by showing the projections of the magnetic field lines in the $x-z$ plane for the D-AD (top) and U-AD (bottom) simulation runs. In can be seen that the field lines form a layer of strongly inclined fields in the upper photosphere and connect at relatively low heights in the D-AD run. The field is mostly weak and horizontal at chromospheric heights. Unlike that, the field is extending into higher layers in the U-AD run and is more aligned with the vertical direction, forming classical flux-tube like expanding structures.

\begin{figure*}[t]
\begin{center}
\includegraphics[width = 16cm]{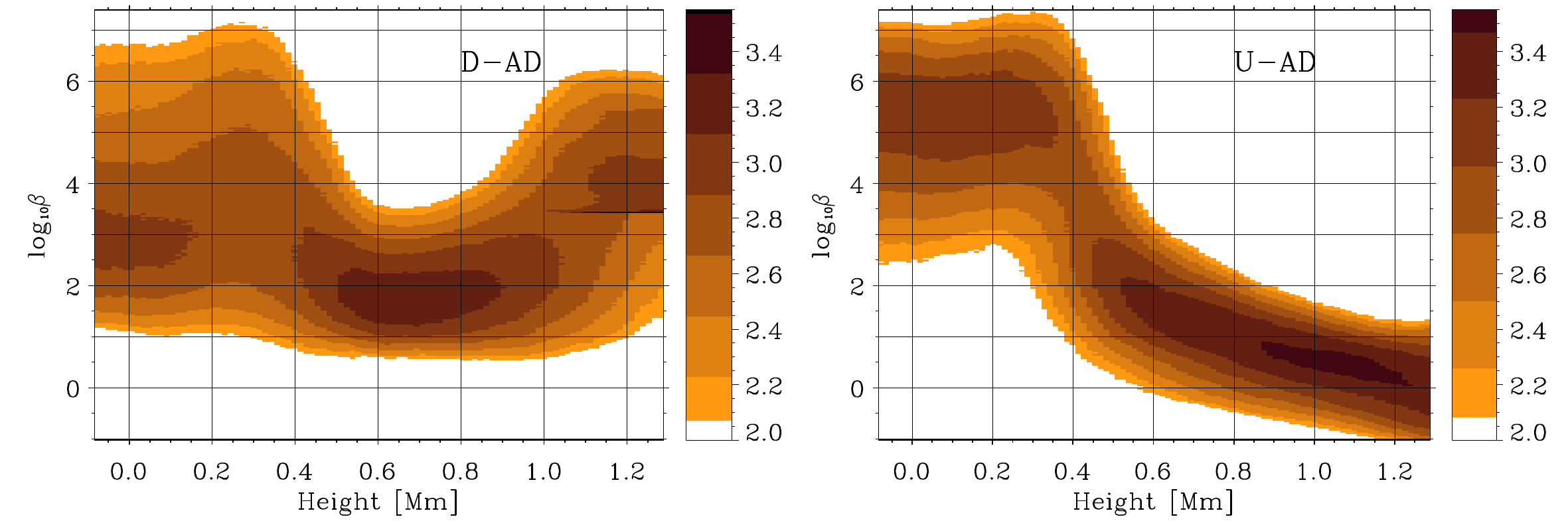}
\end{center}
\caption{Bi-dimensional histogram showing the number of occurrences of a given value of plasma $\beta$ as a function of height in D-AD run (left) and U-AD run (right). Darker colors mean larger probability of occurrence in logaritmic scale, indicated by the color bar.}
\label{fig:beta}
\end{figure*}

\begin{figure*}
\begin{center}
\includegraphics[width = 16cm]{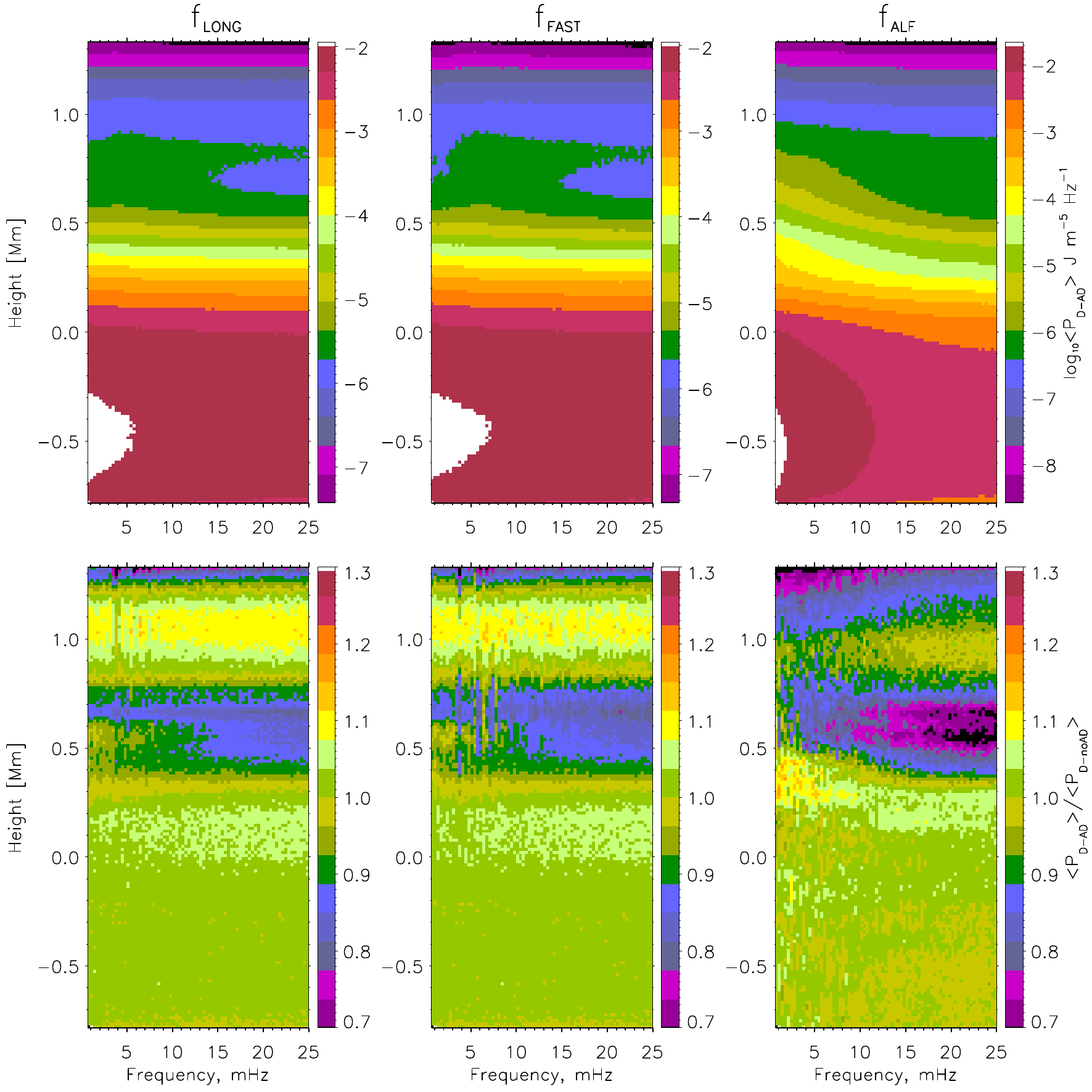}
\end{center}
\caption{Top: power spectra of f$_{\rm long}$ (left), f$_{\rm fast}$ (middle) and f$_{\rm alf}$ (right) as a function of height and frequency for the D-AD run. For better visualization, the maps are scaled with a factor $\rho_0$, the density of the unperturbed atmosphere. Bottom: ratio between D-AD and D-noAD power maps. The power is shown in log10 units of J m$^{-5}$ Hz$^{-1}$. } \label{fig:powertime_battery_v}
\end{figure*}

\begin{figure*}
\begin{center}
\includegraphics[width = 12cm]{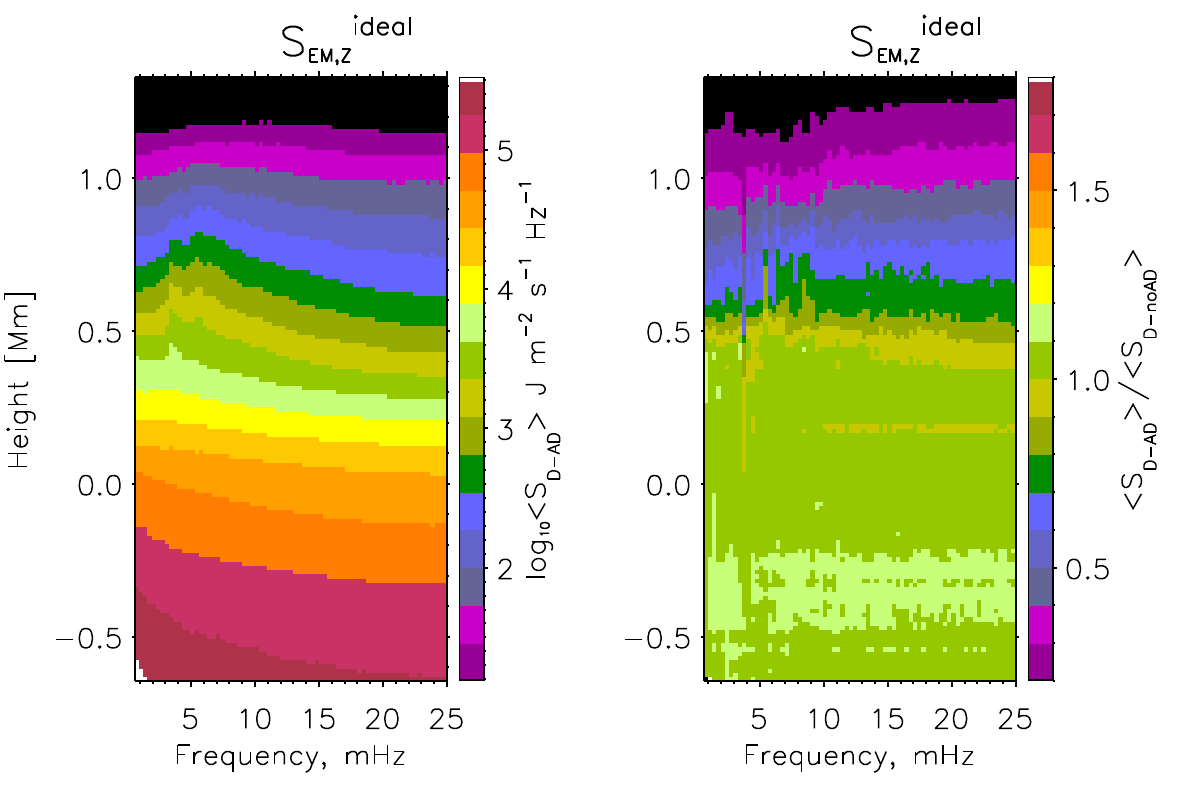}
\end{center}
\caption{Amplitude spectra of the vertical component of the Poynting flux,  $\mathbf{S}_{\rm EM, z}^{\rm ideal}$ (ideal part), defined by Eq. \ref{eq:poynting-ideal}, as a function of height and frequency.  Left: amplitude map for D-AD run; right column: ratio between D-AD and D-noAD amplitude maps. The amplitude is shown in log10 units of J m$^{-2}$ s$^{-1}$ Hz$^{-1}$.   }\label{fig:powertime_battery_fmag}
\end{figure*}

\section{Statistical Fourier analysis}

The patterns of convection and magnetic field distribution developed in the AD and noAD runs are different. Therefore their properties can only be compared in a statistical way. Below we perform a statistical analysis of different aspects of the simulations. We are primarily interested in the properties of waves generated in the simulations, and their dissipation at chromospheric heights. We will also compare the amount of heating and thermal properties of the D and U simulations. 

Convective motions, similar to any turbulent flow, generate waves. The generation of waves in convection simulations have been studied elsewhere \citep{Stein+Nordlund2001, Nordlund+Stein2001}. The high cadence and length of our simulated series allows us to study waves up to high frequencies. The wave properties are intrinsically related to the strength and topology of the magnetic field configuration in the medium where they propagate. Therefore, in order to separate as much as possible different wave modes we base on their physical properties and construct quantities serving as proxies for these modes. In a horizontally homogeneous plasma and in the cold plasma approximation, when no acoustic modes exist, \citet{Cally2017} separated the fast magnetic and the Alfv\'en modes by calculating the divergence and the curl of the wave velocity, correspondingly. This works well because the Alfv\'en mode is incompressible, and the only mode involving compressibility is the fast magnetic mode. In a warm and horizontally structured plasma, one has to deal with at least three wave modes (fast and slow magneto-acoustic and Alfv\'en), keeping in mind that none of them is a pure mode as in a homogeneous plasma. In order to maximize the separation between these modes, we follow the strategy as in  \citet{Cally2017}, and construct the following quantities:
\begin{equation} \label{eq:falf}
\mathrm{f}_{\rm alf} = \hat{e}_\parallel \,\mathbf{\cdot \nabla \times} \mathbf{v},
\end{equation}
\begin{equation}\label{eq:ffast}
\mathrm{f}_{\rm fast} =\mathbf{\nabla\, \cdot} \left(\mathbf{v} -  \hat{e}_\parallel \,v_{\parallel}\right) = \mathbf{\nabla\cdot v}_\perp,
\end{equation}
\begin{equation}\label{eq:flong}
\mathrm{f}_{\rm long} = \hat{e}_\parallel \,\mathbf{\cdot \nabla} \left(\mathbf{v} \cdot \hat{e_\parallel} \right)  = \vec{\nabla}_\parallel v_{\parallel}.
\end{equation}
Notice that the quantities defined by equations (\ref{eq:falf}--\ref{eq:flong}) are not velocities but their derivatives. Here, $v_{\parallel} =  \hat{e}_\parallel \cdot \mathbf{v}$ and $\mathbf{v}_\perp=\mathbf{v} -  \hat{e}_\parallel \,v_{\parallel}$ are the velocity components parallel and perpendicular to the magnetic field, and $ \hat{e}_\parallel$ is the field-aligned unit vector. A similar approach was also used by \citet{Przybylski+etal2017} in a simpler situation for the study of waves in an isolated expanding flux tube, showing good results. Independently of the plasma $\beta$, $\mathrm{f}_{\rm alf}$ gives an incompressible perturbation propagating along the field lines and separates the Alfv\'en waves. The quantity $\mathrm{f}_{\rm long}$ gives the compressible perturbation propagating along the field lines, and separates the slow mode (essentially acoustic) waves in $\beta<1$. The remaining quantity, $\mathrm{f}_{\rm fast}$ gives the compressible perturbation in the direction perpendicular to the field lines, separating the fast mode (essentially magnetic) waves for $\beta<1$. 

The plasma $\beta$ in the simulations vary strongly as a function of the spatial coordinates and time, depending on the evolution of magnetic structures. The bi-dimensional histograms of the plasma $\beta$ as a function of height are given in Figure \ref{fig:beta}. It can be seen that, for the weakest fields in the $D$ runs, the lowest values of $\beta$ are achieved between 600 and 800 km, and they are almost always above unity. In the $U$ runs, progressively larger part of the domain contains plasma with $\beta<1$ starting from approximately 600 km height.

\begin{figure*}
\begin{center}
\includegraphics[width = 16cm]{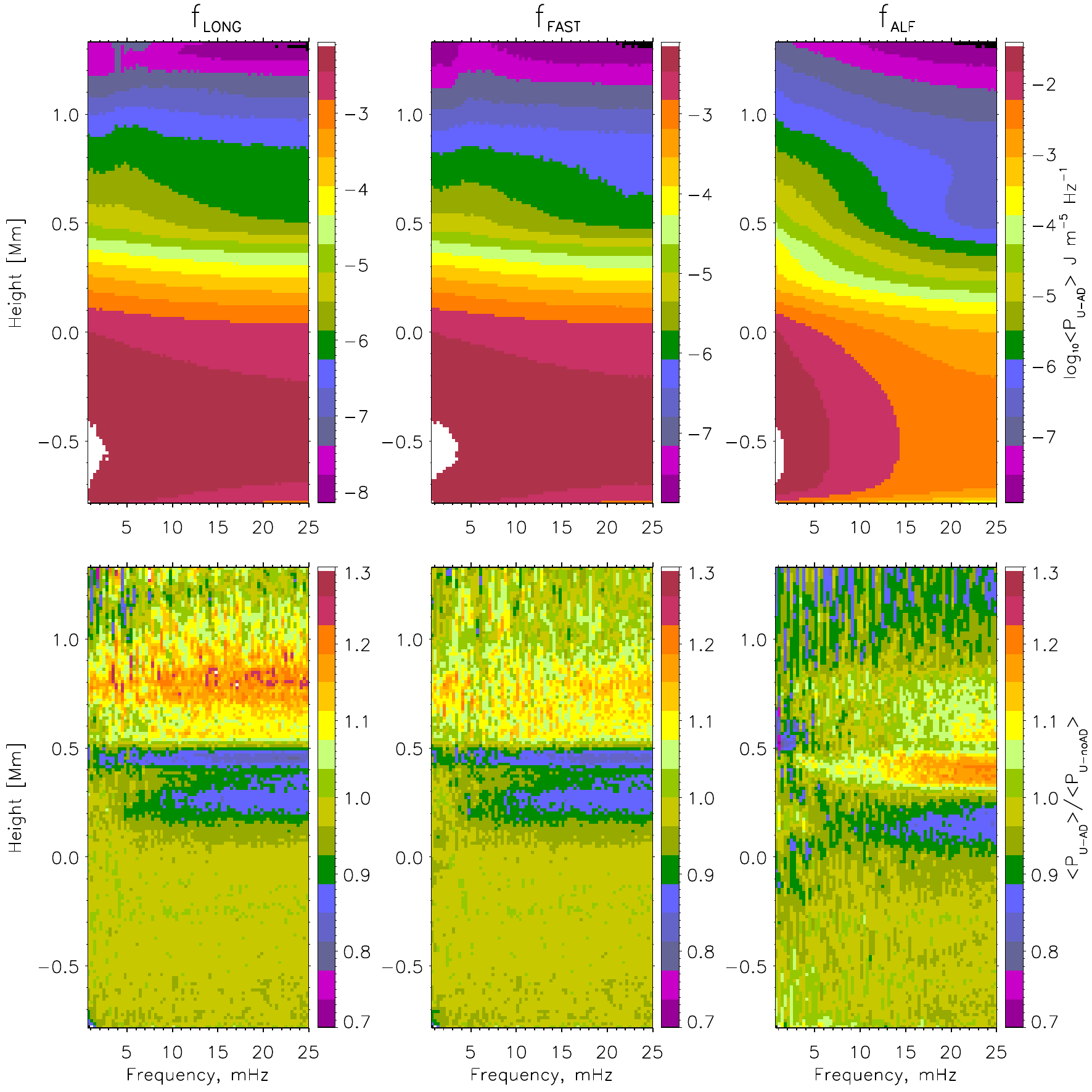}
\end{center}
\caption{Top: power spectra of f$_{\rm long}$ (left), f$_{\rm fast}$ (middle) and f$_{\rm alf}$ (right) as a function of height and frequency for the U-AD run. For better visualization, the maps are scaled with a factor $\rho_0$, the density of the unperturbed atmosphere. Bottom: ratio between U-AD and U-noAD power maps. The power is shown in log10 units of J m$^{-5}$ Hz$^{-1}$. }\label{fig:powertime_10G_v}
\end{figure*}

We constructed power maps of f$_{\rm alf}$, f$_{\rm fast}$, and f$_{\rm long}$, as a function of height and frequency. For that, we computed the temporal power spectra of these quantities at each spatial point and at every height. The spectra were then averaged over the two horizontal directions. These power maps are displayed in Figures \ref{fig:powertime_battery_v} and \ref{fig:powertime_10G_v} for D and U simulations, correspondingly.

We also consider the Poynting flux reaching the upper levels of the simulation domain. The total energy conservation equation can be rewritten in terms of the electro-magnetic Poynting flux as follows:

\begin{equation}
\frac{\partial e_1}{\partial t} + \vec{\nabla}\cdot\left(\mathbf{v} \left( p + e - \frac{|\mathbf{B}^2|}{2\mu_0}\right)  + \frac{\mathbf{E}\times\mathbf{B}}{\mu_0} +  \mathbf{F}_R \right) = \rho\mathbf{v}\cdot\vec{g}, 
\end{equation}
where we have removed the small artificial diffusivity term of Eq. \ref{eq:energy}, and have used the expression for the radiative flux, $Q_R=-\vec{\nabla}\mathbf{F}_R$. The electromagnetic Poynting flux is defined as: 
\begin{equation} \label{eq:poynting}
\mathbf{S}_{\rm EM}=\frac{\mathbf{E}\times\mathbf{B}}{\mu_0}= -\frac{\mathbf{B} \times (\eta_A \mathbf{J_\perp}) }{\mu_0}   - \frac{\mathbf{\nabla}p_e \times \mathbf{B}} {e n_e \mu_0}   -\frac{(\mathbf{v}\times\mathbf{B})\times\mathbf{B}}{\mu_0}  
\end{equation}

The contribution from the battery term to $\mathbf{S}_{\rm EM}$ is expected to be small, and the main difference between simulations with/without AD is expected to come from the presence of the ambipolar term, the first term on the right hand side of Eq. \ref{eq:poynting}. Therefore, we neglect the contribution of the battery term and split $\mathbf{S}_{\rm EM}$ into ideal and ambipolar parts as:
\begin{equation} 
\mathbf{S}_{\rm EM}=\mathbf{S}_{\rm EM}^{AD} + \mathbf{S}_{\rm EM}^{\rm ideal},
\end{equation}
with
\begin{equation} 
\mathbf{S}_{\rm EM}^{AD} =-\frac{\mathbf{B} \times (\eta_A \mathbf{J_\perp}) }{\mu_0},   
\end{equation}
and
\begin{equation} \label{eq:poynting-ideal}
\mathbf{S}_{\rm EM}^{\rm ideal}=-\frac{(\mathbf{v}\times\mathbf{B})\times\mathbf{B}}{\mu_0}. 
\end{equation}

\begin{figure*}
\begin{center}
\includegraphics[width = 12cm]{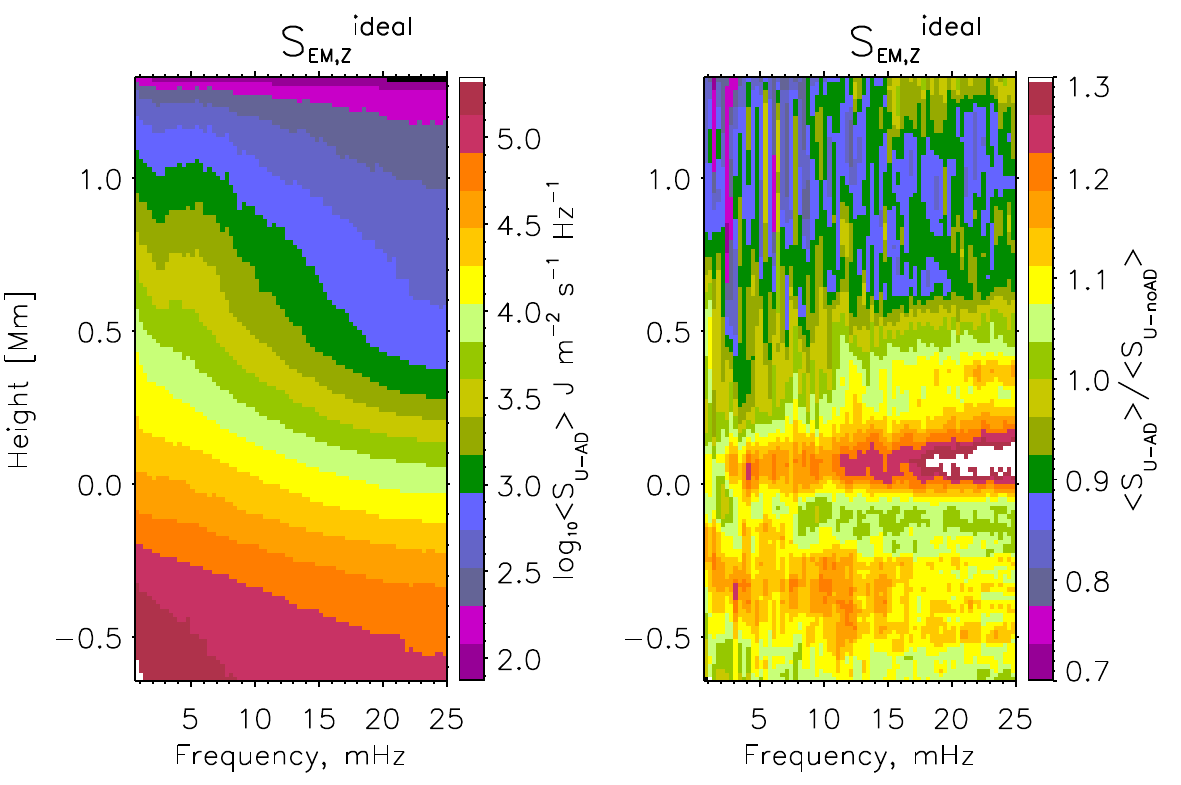}
\end{center}
\caption{Amplitude spectra of the vertical component of the Poynting flux,  $\mathbf{S}_{\rm EM, z}^{\rm ideal}$ (ideal part), defined by Eq. \ref{eq:poynting-ideal}, as a function of height and frequency.  Left: amplitude map for U-AD run; right column: ratio between U-AD and U-noAD amplitude maps. The amplitude is shown in log10 units of J m$^{-2}$ s$^{-1}$ Hz$^{-1}$.}\label{fig:powertime_10G_fmag}
\end{figure*}

The Fourier amplitude maps of the vertical component of the ideal part of the Poynting flux,  $\mathbf{S}_{\rm EM, z}^{\rm ideal}$, are displayed in Figures \ref{fig:powertime_battery_fmag} and \ref{fig:powertime_10G_fmag}, for the D and U simulations, correspondingly.

Before proceeding with the description of the results, it is interesting to check which velocities create the ideal part of the Poynting flux in the U and D simulations. For that we split  $\mathbf{S}_{\rm EM}^{\rm ideal}$ into ``horizontal'' and ``vertical'' parts, following  \citet{Shelyag2012},
\begin{eqnarray}
S_{\rm EM}^{\rm ideal, vert}&=&v_z\cdot(B_x^2 + B_y^2)/\mu_0; \\
S_{\rm EM}^{\rm ideal, hor}&=&B_z\cdot(B_x v_x + B_y v_y)/\mu_0.
\end{eqnarray}
The ``horizontal'' part corresponds to horizontal motions along vertical flux tubes, while the ``vertical'' part corresponds to horizontal magnetic field perturbations transported by vertical plasma motions. In their paper, \citet{Shelyag2012} conclude that torsional motions in the magnetic flux tubes produce most of the ideal Poynting flux (i.e. $S_{\rm EM}^{\rm ideal, hor} > S_{\rm EM}^{\rm ideal, vert}$, while earlier \citet{Steiner2008} claimed that the largest part of this flux is due to the vertical motions and magnetic field advection, i.e. the opposite is true. In our case, we get that for D simulations, $S_{\rm EM}^{\rm ideal, vert}$ is about an order of magnitude smaller than  $S_{\rm EM}^{\rm ideal, hor}$, similar to the results by \citet{Shelyag2012}. In the U simulations the situation is not so straightforward. We obtain that $S_{\rm EM}^{\rm ideal, hor} < S_{\rm EM}^{\rm ideal, vert}$ below the surface and they are similar above the surface. This is an interesting behavior and will need a future detailed study. 

\subsection{D-runs}

Figure \ref{fig:powertime_battery_v} shows the power maps of the quantities defined by equations (\ref{eq:falf}--\ref{eq:flong}) for the D-AD run (top), and the ratio between the D-AD and D-noAD power maps (bottom). We do not show the maps for the D-noAD case, since due to the logarithmic change of the power with height, the differences with the D-AD maps cannot be appreciated well. 

The power maps show the spectrum characteristic for convective motions below the photosphere (negative heights). There is hardly any difference in power between the D-AD and D-noAD cases for heights below 0 km, see the bottom panels of Fig. \ref{fig:powertime_battery_v} (green colors marking values close to 1). The power distributions for $\mathrm{f}_{\rm long}$, $\mathrm{f}_{\rm fast}$, and $\mathrm{f}_{\rm alf}$ show most differences in the atmospheric layers, where the magnetic field has an increasing influence. 


The power maps of the $\mathrm{f}_{\rm long}$ and $\mathrm{f}_{\rm fast}$ look rather similar to each other. This is not surprising since both quantities select compressible waves, and the values of the plasma $\beta$ in the D-AD run are not low enough to provide anisotropy in the compressible wave propagation along and across the magnetic field. The power of $\mathrm{f}_{\rm long}$ and $\mathrm{f}_{\rm fast}$ in the atmospheric layers reveals a spectrum that is typical for the solar 5-minute oscillations with a broad maximum between 2 and 10 mHz. The length of the simulated series (40 min) does not allow to resolve individual power peaks well. Still, the power maps ratios (left and middle bottom panels) show discrete vertical strips pattern at frequencies between 2 and 10 mHz, reflecting slight differences between power peaks locations in the D-AD and D-noAD runs, probably caused by differences in their average temperature structures.
 
The spectrum of the $\mathrm{f}_{\rm alf}$ quantity (top right panel of Fig. \ref{fig:powertime_battery_v}) is different from that of $\mathrm{f}_{\rm long}$ and $\mathrm{f}_{\rm fast}$, showing more power at low frequencies. The dominance of Alfv\'en waves (selected by the $\mathrm{f}_{\rm alf}$ quantity) in the middle photosphere above 500 km can be explained by the presence of the layer with strongly inclined fields in the simulations at those heights (see Figs. \ref{fig:bmean} and \ref{fig:tcut}), which produce favorable conditions for the fast to Alfv\'en mode transformation \citep{Cally+Goossens2008, Khomenko+Cally2012}. 

The ratio between the power maps of the $\mathrm{f}$-quantities, displayed in the bottom panels of Fig. \ref{fig:powertime_battery_v}, tells us about the differences in the amount of wave energy present at a given height and at a given frequency, depending on whether the ambipolar term is operating or not. The power ratio maps remarkably show the presence of power depression layers, where the power of waves in the D-AD simulations is decreased compared to that of the D-noAD simulations, see the violet-purple-black strips at the bottom panels of Fig. \ref{fig:powertime_battery_v}. This decrease reaches up to 30-40 \% for the $\mathrm{f}_{\rm alf}$ quantity. The location and extension of the power depression layers depends on the $\mathrm{f}$-quantity and on the wave frequency. It is interesting to note that the depression is not a smooth function of frequency in none of the cases.  Both $\mathrm{f}_{\rm long}$ and $\mathrm{f}_{\rm fast}$ have a power depression layer located between 400 and 800 km, roughly coinciding with the heights where strongly inclined fields exist in the D-AD and D-noAD simulations (see Figs. \ref{fig:bmean} and \ref{fig:tcut}). The height coverage (width) of the depression layer increases with increasing frequency. The strength of the depression also increases with frequency. There is a secondary power depression layer located higher up above 1.2 Mm as well. 

It is remarkable that the power depression is significantly stronger for the $\mathrm{f}_{\rm alf}$ quantity, representing incompressible perturbations. In the power depression layer located between 400 and 800 km, the amount of depression increases toward the high frequencies. In the secondary power depression layer located above 1 Mm, this behavior reverses, and the amount of depression decreases with frequency, being largest for incompressible waves with the lowest frequencies. 

There can be several alternative explanations for the power distributions. On the one hand, the structure of the field can be different in the D-AD and D-noAD simulations due to the proper effect of the ambipolar diffusion on the field. In the AD simulations the field may reach higher heights by diffusing through the neutral chromospheric layers, as in \citet{Leake+Arber2006, Arber2007}.  This may displace the regions with maximum wave power at a given frequency to a different layer. On the other hand, the change of the average field structure can affect the mode transformation process, making the relative weight between the various wave modes change with height and frequency in a different way for the D-AD and D-noAD simulations. The efficiency of the mode transformation could also be affected directly by the presence of the ambipolar diffusion, see the recent study by \citet{Cally+Khomenko2018}.

Apart from the possible change in the average field structure and efficiency of the mode transformation process, real physical absorption and wave power conversion into heat can be taking place. Wave absorption due to dissipation of the ambipolar currents was studied recently by \citet{Shelyag+etal2016} using idealized simulations of wave propagation in isolated flux tubes. There, waves at a single frequency of 25 mHz were studied. While the structure of the field was significantly different in the simulations by \citet{Shelyag+etal2016}, it is interesting that a similar effect is now confirmed in a more complex and realistic situation. Additionally, a similar non-smooth behavior of the wave power depression with frequency was found by \citet{Przybylski+etal2017} in a work continuing the initial results by  \citet{Shelyag+etal2016}. Fully understanding these effects would require more idealized simulations to be done. In particular, unlike in idealized cases, here we observe a significant power depression for the longitudinal component of the velocity. This peculiarity may have to do with the particular field structure in the current simulation run (see next section). 

The Poynting flux tells us about the propagation of wave energy, and not only about its amount at a given height. To check if there is physical absorption of the wave energy, Figure \ref{fig:powertime_battery_fmag} shows amplitude maps of the vertical component of the ideal part of the Poynting flux vector, Eq. \ref{eq:poynting-ideal}. Similar to the velocity power maps, the Poynting flux in the deep layers is essentially the same in the D-AD and D-noAD runs (see the green colors of the right panel). It also shows the presence of a broad power maximum for frequencies characteristic for solar waves, between 3 and 10 mHz above the surface. The most prominent feature, however, is the strong decrease of the $\mathbf{S}_{\rm EM}$ in the D-AD simulation compared to the D-noAD one in the upper layers. The absorption of the flux reaches up to 80-90\% in the uppermost part of the simulation domain, being larger for lower frequencies. The amount of Poynting flux absorption coefficient agrees well with that reported by \citet{Shelyag+etal2016} from idealized simulations. 

\begin{figure*}
\begin{center}
\includegraphics[width = 18cm]{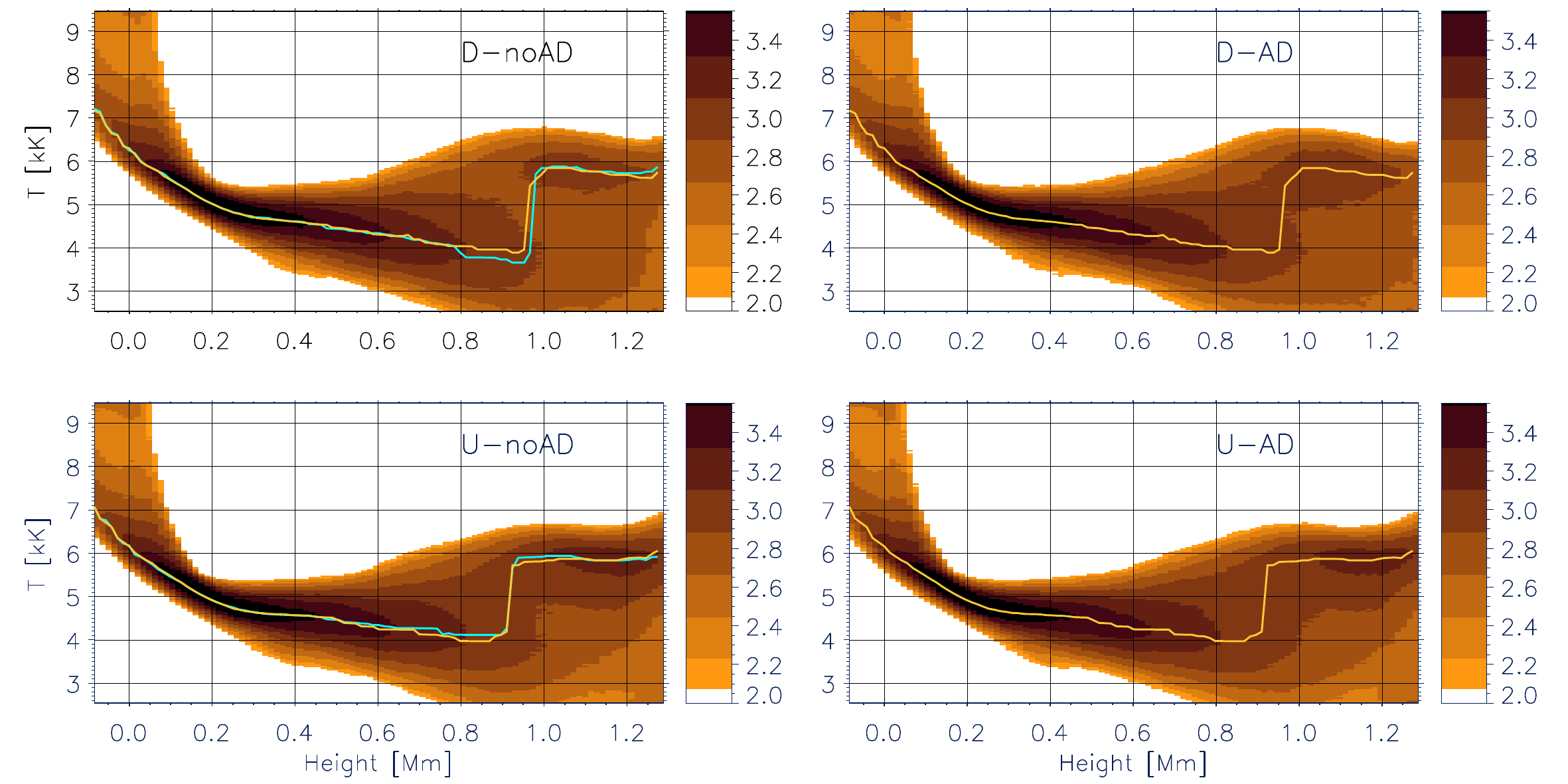}
\end{center}
\caption{Bi-dimensional histograms showing the logarithm of the number of points with a given value of temperature at a given height. The darker colors mean larger number of points, indicated at the bar in log$_{10}$ scale. Upper left: D-noAD run; upper right: D-AD run; bottom left: U-noAD run; bottom right: U-AD run. Yellow/light blue lines follow the maximum value of the distribution at each height for AD/noAD runs, correspondingly.  }\label{fig:temperature}
\end{figure*}

\subsection{U-runs}

The magnetic field is mostly unipolar in the U runs. Due to the granular motion, most of the magnetic field is moved into the intergranular lanes where it forms nearly vertical expanding flux tubes with a photospheric strength close to kG values, see Figs. \ref{fig:bexample} and \ref{fig:tcut}. The average field strength in the atmospheric layers is larger in the U simulations compared to the D simulations, see Fig. \ref{fig:bmean}. Due to this more regular and generally ``less chaotic'' field topology, the combination of wave modes developed in the U simulations is different to that of the D simulations. The power maps of the f-quantities, given in Figure \ref{fig:powertime_10G_v}, reflect these changes.

Similar to the D runs, the power maps of the $\mathrm{f}_{\rm long}$ and $\mathrm{f}_{\rm fast}$ quantities show a broad power peak in the atmospheric layers between 2 and 10 mHz, characteristic for solar 5-min oscillations (upper left and middle panels). The spectrum of the $\mathrm{f}_{\rm long}$ (upper right) provides a smoother decrease of the power with frequency, with enhanced presence of the low-frequency waves. 

The most significant difference, however is in the power depression maps shown at the bottom panels of Fig. \ref{fig:powertime_10G_v}. Unlike the D-runs, the amount of power depression is significantly less (10-15\%) and there is no secondary depression band above 1 Mm for the  $\mathrm{f}_{\rm long}$ and $\mathrm{f}_{\rm fast}$ quantities.  In fact, the power of compressible waves, selected by $\mathrm{f}_{\rm long}$ and $\mathrm{f}_{\rm fast}$, in the U-AD simulations is up to 20-25\% higher above 0.5 Mm compared to the U-noAD case. The amount of power depression is still a function of frequency and it is increasing for higher frequencies. 

The power ratio maps for the incompressible waves, selected by $\mathrm{f}_{\rm alf}$ (bottom right) show a secondary extended power depression band above 800 km, where the maximum value of depression is still below 15\%. There is about 20\% power enhancement for highest frequencies at heights around 400 km.

Since the ambipolar diffusion coefficient, Eq. \ref{eq:etaa}, is proportional to the magnetic field strength squared, one could intuitively expect that the wave dissipation should be larger in models with average larger field strengths, as we have in the U runs. However, our simulations show the opposite results. This brings us to the conclusion that the efficiency of the wave power absorption via AD mechanism is intimately related to the magnetic field structure. The eventual efficiency of the energy transfer will not only depend on the strength of $\eta_A$, but also on the strength and abundance of perpendicular currents to dissipate. The more chaotic and bi-polar field structures developed in the D simulations apparently provide a better source of perpendicular currents than the more regular and unipolar fields in the U runs.

The amplitude maps of the Poynting flux, shown in Figure \ref{fig:powertime_10G_fmag}, confirm this conclusion. We observe that the Poynting flux has less contribution at high frequencies in the U runs, compared to the D runs. There is a layer above 0.5 Mm where the absorption due to the AD mechanism is present. However, the amount of absorption is significantly less in the U case, with up to 30\%, compared to 90\% in the D runs. There is also a layer around 0--0-2 Mm where the Poynting flux in the U-AD case is about 50\% larger compared to the U-noAD case (orange-red-white strip). The origin of this strip should be investigated further. 

It is important to underline the difference between the power ratio maps of the f-quantities (bottom panels of Fig. \ref{fig:powertime_10G_v}) and the amplitude ratio map of the Poynting flux (right panel of Fig. \ref{fig:powertime_10G_fmag}). The locations of the strips where the amount of the available kinetic energy is less in the U-AD runs compared to U-noAD runs do not coincide with the location of the strip where the Poynting flux is absorbed, except for the upper depression band for $\mathrm{f}_{\rm alf}$. This happens because the quantities studied are fundamentally different. While the power maps in Fig. \ref{fig:powertime_10G_v} indicate the amount of kinetic energy available, the Poynting flux indicates the amount of electromagnetic energy that propagates through the region. For example, the wave energy cannot propagate for waves that are evanescent or standing or trapped by some mechanism; even if their $\rho$v$^2$ is large, their energy flux might be zero. The Poynting flux absorption maps, as in  Figure \ref{fig:powertime_10G_fmag}, indicate the amount of wave electromagnetic energy physically absorbed in a given layer. 

\begin{figure*}
\begin{center}
\includegraphics[width = 18cm]{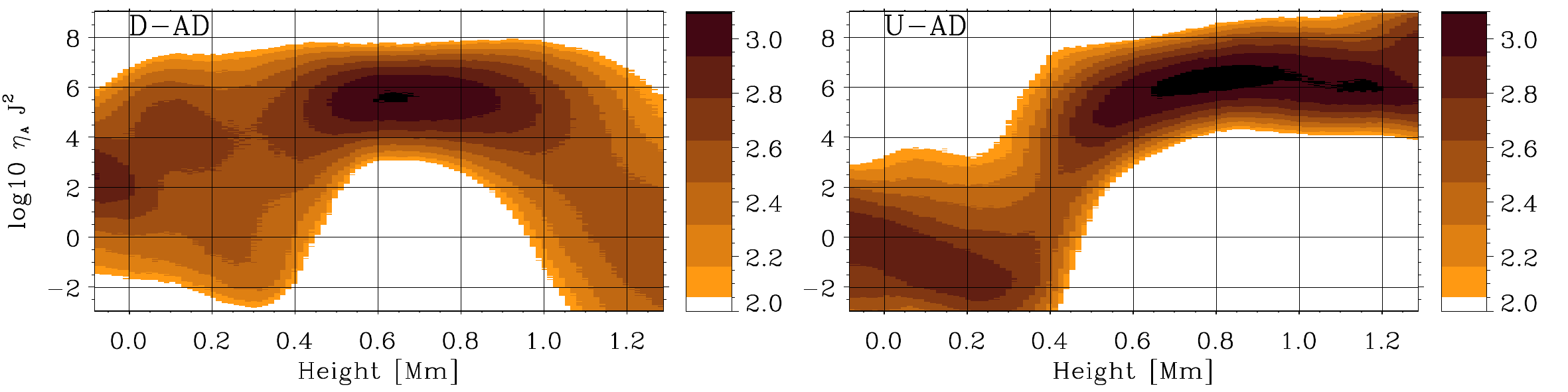}
\end{center}
\caption{Bi-dimensional histograms showing the logarithm of the number of points with a given value of the heating term, $\eta_AJ_{\perp}^2$ as a function of height. Darker colors mean larger number of points, indicated by the color bar in log$_{10}$ scale. Left: D-AD run; right: U-AD run. Notice the difference in the heights with a maximum heating term between both cases.  }\label{fig:heat}
\end{figure*}

\begin{figure*}
\begin{center}
\includegraphics[width = 16cm]{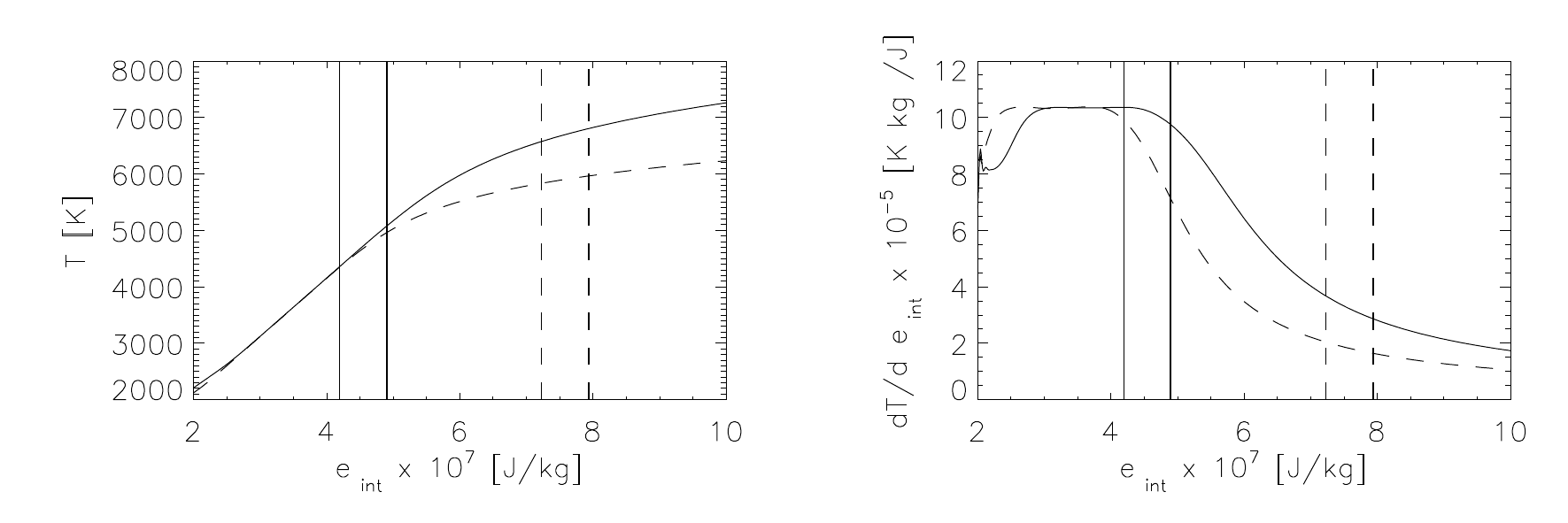}
\end{center}
\caption{Left panel: cuts through EOS table showing $T(e_{\rm int})$ for fixed $\rho$ values; solid line is for densities corresponding to 600 km and dashed line is for densities corresponding to 1100 km. Vertical solid and dashed thin lines mark the values of $e_{\rm int}$ corresponding to 600 and 1100 km. Vertical solid and dashed thick lines are $e_{\rm int}+\eta_A J^2$, taking the same value for $\eta_A J^2=10^7$, as per Figure \ref{fig:heat}. Right panel: same but for the gradient $dT/d e_{\rm int}$. }
\label{fig:eos}
\end{figure*}

\section{Heating properties of the simulations}

The absorption of Poynting flux allows to convert wave energy into heat and should eventually be reflected on the average temperature structure in the simulations. Since we only have a simplified treatment of radiative transfer, using LTE in the chromospheric layers, we can only extract conclusions about the average temperature increase by comparing statistically the AD and noAD cases. Figure \ref{fig:temperature} shows the bi-dimensional histograms of the temperature distribution as a function of height in the four simulation runs. The histograms include all spatial points and the whole temporal series duration. The figure only shows the height range from the photosphere upwards where the difference due to the presence of the AD effect is expected to be important. While in general the temperature stratifications between the AD and noAD cases seem rather similar, there are also significant differences. The upper right panel of Fig. \ref{fig:temperature} demonstrates that there are more locations with higher chromospheric temperatures (heights between 0.8 and 1 Mm). The difference between AD and noAD cases is better appreciated by comparing the yellow and light blue curves that follow the maximum value of the distribution at every height. It can be seen that at heights between 0.8 and 1 Mm in the D run (upper panels), the maximum of the distribution is shifted toward 200-300 K higher values for the AD case. On average, the temperatures in the D run are about 50 K larger in the AD case at these heights. However, it should be noted that the temperature distributions at all heights are wide and have a complex shape, so that mean values are not the best quantities to characterize such distributions.

The average temperatures are still slightly larger in the U-AD case compared to the U-noAD case (low panels of Fig. \ref{fig:temperature}), but the effect is significantly smaller than in D cases.  In fact, the maximum values of the distribution are larger in the U-noAD case, showing just the opposite effect to the D cases. This can be a consequence of the lower amount of absorbed Poynting flux in the U simulations compared to the D simulations. 

\begin{figure*}
\begin{center}
\includegraphics[width = 19cm]{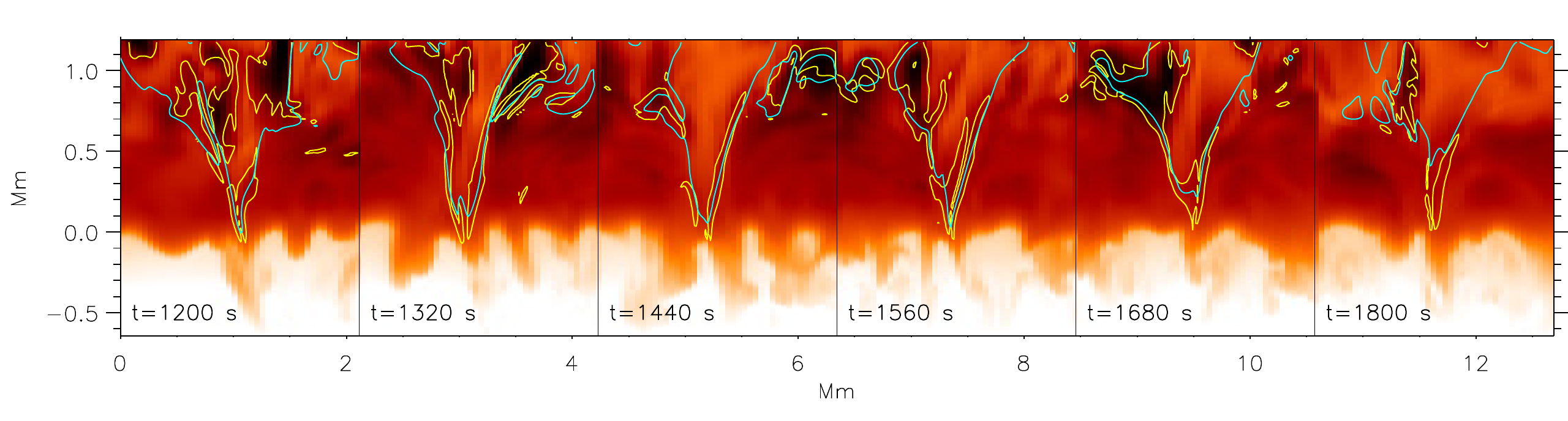}
\end{center}
\caption{Time series of vertical cuts showing the temperature structure (red colors), contours of log$_{\rm 10}( \eta_A J_{\perp}^2)=8$ (yellow lines), and contours of log$_{\rm 10}\beta=0$ (plasma $\beta$, blue lines) in the U-AD simulations. Darker red colors mean low temperatures. The time of the snapshots is indicated in the figure.}\label{fig:te_heat_beta_10G}
\end{figure*}
\begin{figure*}
\begin{center}
\includegraphics[width = 19cm]{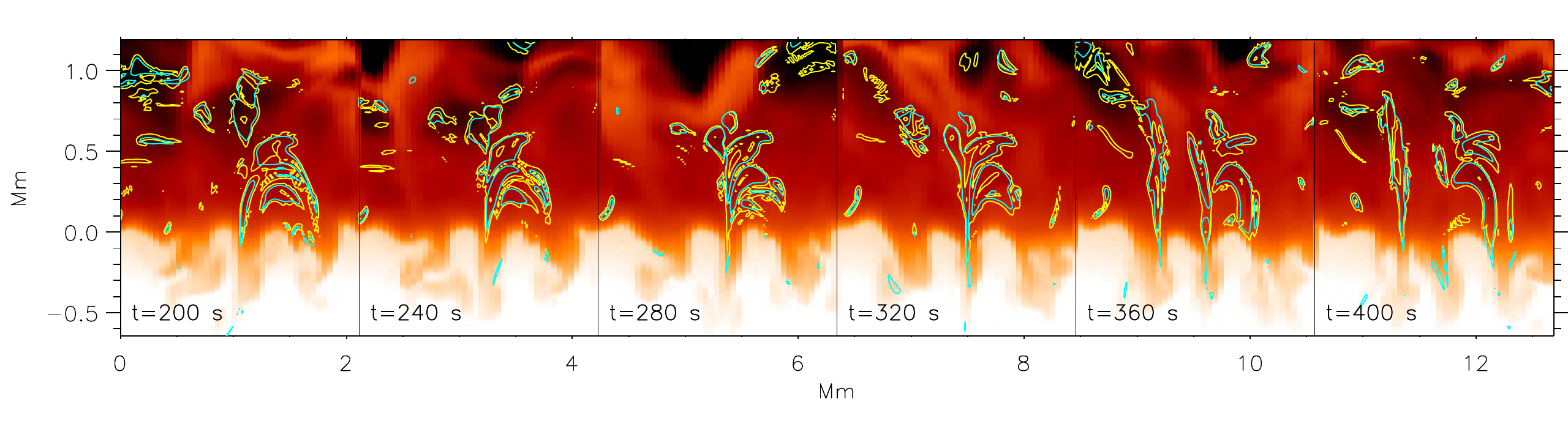}
\end{center}
\caption{Time series of vertical cuts showing the temperature structure (red colors), contours of log$_{\rm 10}( \eta_A J_{\perp}^2)=7$ (yellow lines), and contours of log$_{\rm 10}\beta=1$ (plasma $\beta$, blue lines) in the D-AD simulations. Darker red colors mean low temperatures. The time of the snapshots is indicated in the figure.} \label{fig:te_heat_beta_battery}
\end{figure*}

Figure \ref{fig:heat} shows the histograms of the distribution of the heating term, $\eta_AJ_{\perp}^2$,  as a function of height in the D-AD (left) and U-AD (right) simulations. The energy conservation equation, Eq. \ref{eq:energy}, can be rewritten in terms of conservation of internal energy only, by removing  the kinetic and magnetic energy parts using the momentum and induction equations:
\begin{equation}
\frac{\partial e_1}{\partial t} + \vec{\nabla}\cdot( \vec{v}e) +p\nabla\cdot \vec{v}  = \eta_AJ_{\perp}^2 + Q_R
\end{equation}
The right hand side of this equation contains the Ohmic heating term related to the ambipolar diffusion. This term gives the efficiency of conversion between the magnetic and thermal energies, see \citet{Khomenko+Collados2012}, and is the quantity shown in Figure \ref{fig:heat}. The efficiency of the energy conversion depends on the magnitude of the ambipolar diffusion coefficients, but also on the strength of the perpendicular currents, $J_{\perp}$. While the first one is essentially a function of temperature and density (through the collisional frequency), the latter is related to the magnetic field structure. Since $\eta_A$ is inversely proportional to the collisional frequency, it gets larger in those areas with lower density. The $J_{\perp}$ gets larger for a more complex field structure with mixed polarities at small spatial scales. 

The comparison between the left and right panels of Fig. \ref{fig:heat} demonstrates that the heights where $\eta_AJ_{\perp}^2$ reaches the largest values are significantly different in the D-AD and U-AD cases. In the D-AD case, large values of the heating term are reached over most of the photosphere up to the chromosphere. The heating term is decreasing strongly above about 1.1 Mm heights since most horizontal fields in the D simulations do not reach so high (their strength significantly decreases in the chromosphere, see Figure  \ref{fig:tcut}). Unlike that, the strong heating only starts above about 0.6 Mm in the U-AD case, but it continues over the whole chromosphere without significant decrease. The distributions of the heating term are consistent with the results of the Poynting flux absorption in Figs. \ref{fig:powertime_battery_fmag} and \ref{fig:powertime_10G_fmag}. There one can see that a significant absorption starts at lower heights in the D-AD case (above 0.5 Mm) compared to the U-AD case (above 0.8 Mm), which roughly coincides with the heights where the largest number of points has maximum values of the $\eta_AJ_{\perp}^2$ in both simulations.

Therefore, the conclusion that might be drawn is that, for producing an efficient heating by the AD mechanism, it is more important to have a complex mixed-polarity structuring of the field as that present in quiet Sun regions. In that case, due to the large amount of available currents, the efficient action of the heating term starts at lower heights and facilitates the eventual absorption of a larger amount of Poynting flux to give rise to a larger temperature increase compared to more regular and unipolar areas, even if structures (flux tubes) with larger strength are present in the latter. This conclusion is important, since the D-AD case may be considered as representative of local dynamo quiet Sun fields occupying most of the the Sun's surface at any time. The heating due to AD effects may be responsible for the average temperature increase in the relatively quiet Sun. The 10 G unipolar case shows a lower average Poynting flux absorption and a smaller temperature increase due to this effect, since less currents are present in this case and the structures are formed at larger scales implying slower time scales for heating.

\begin{figure}
\begin{center}
\includegraphics[width = 4.2cm]{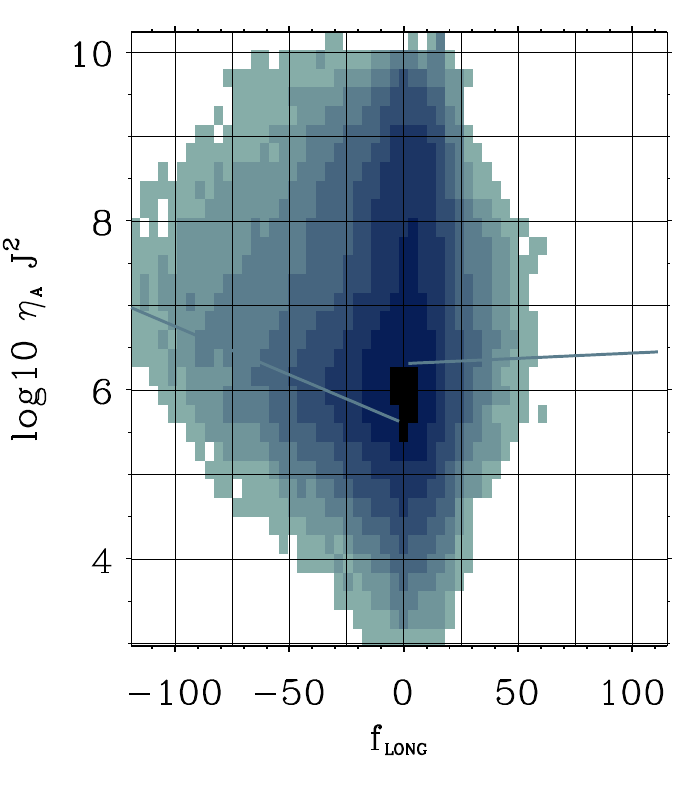}
\includegraphics[width = 4.2cm]{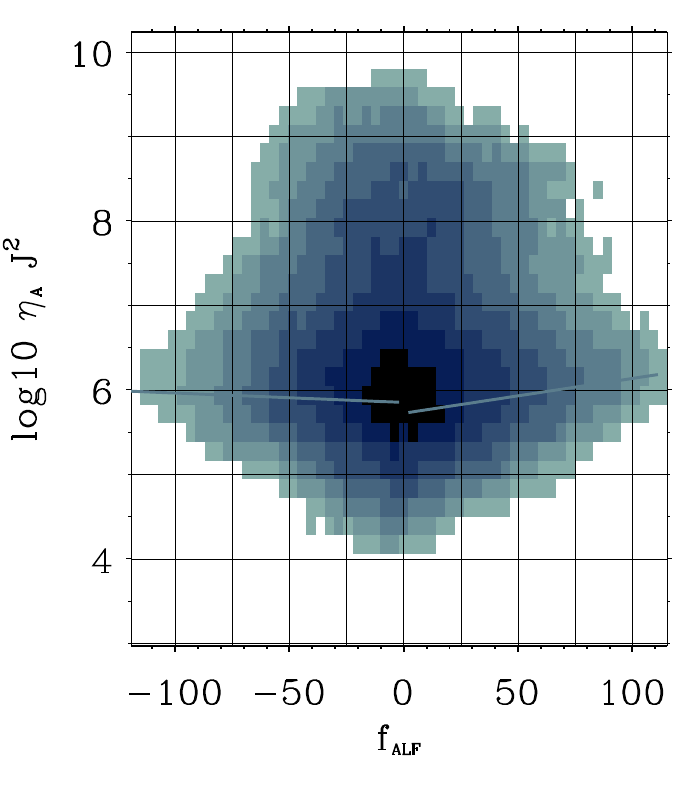}
\end{center}
\caption{Bi-dimensional histograms showing the correlation between the heating term and f$_{\rm long}$ (left) and f$_{\rm alf}$ (right), for the U-AD runs.  The color scale indicates the log$_{10}$ of the number of points with a given combination of values at the axes. Inclined solid lines represent the linear fit to the distribution, done separately for positive and negative values.}\label{fig:mach_10G}
\end{figure}
\begin{figure}
\begin{center}
\includegraphics[width = 4.2cm]{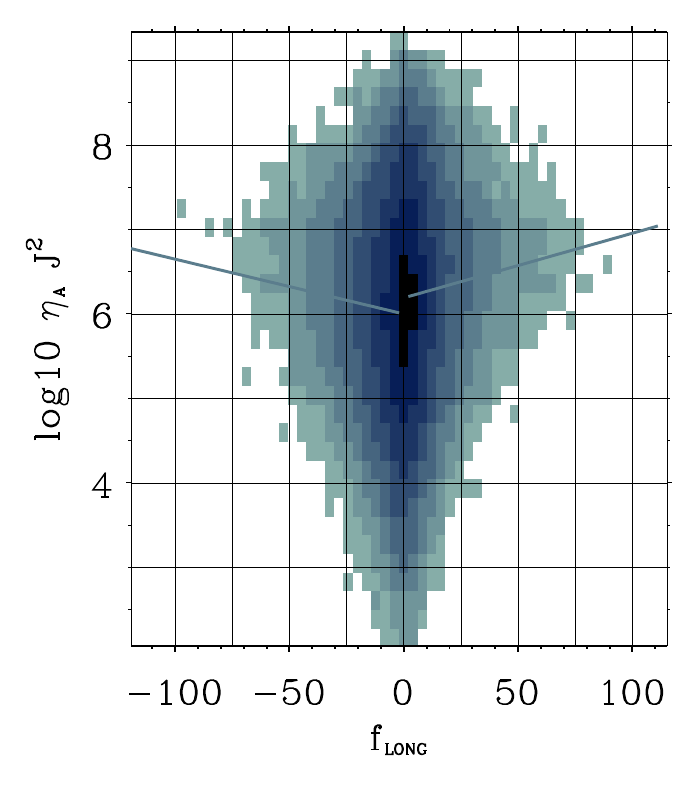}
\includegraphics[width = 4.2cm]{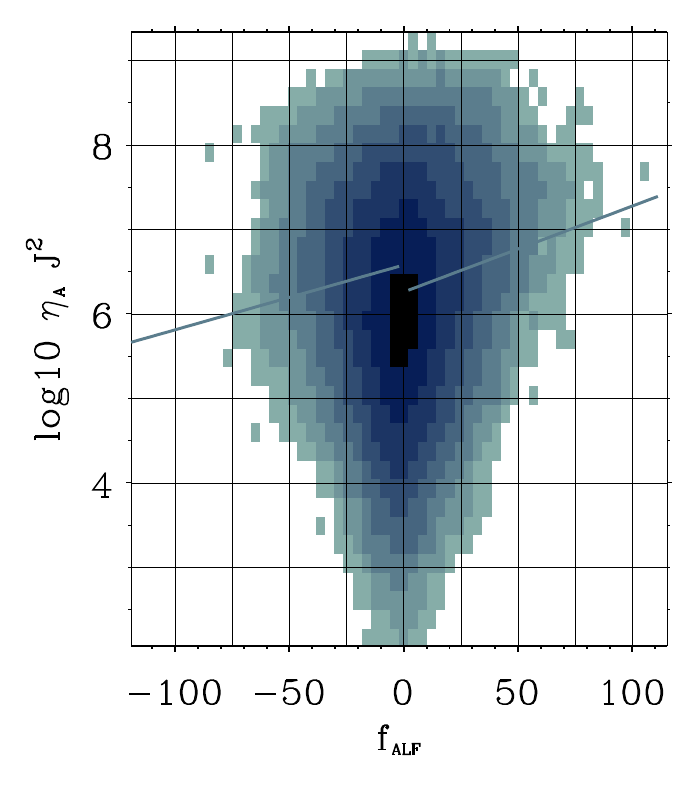}
\end{center}
\caption{Same as  Fig. \ref{fig:mach_10G}, but  for D-AD runs.}\label{fig:mach_bat}
\end{figure}

\begin{figure*}
\begin{center}
\includegraphics[width = 12cm]{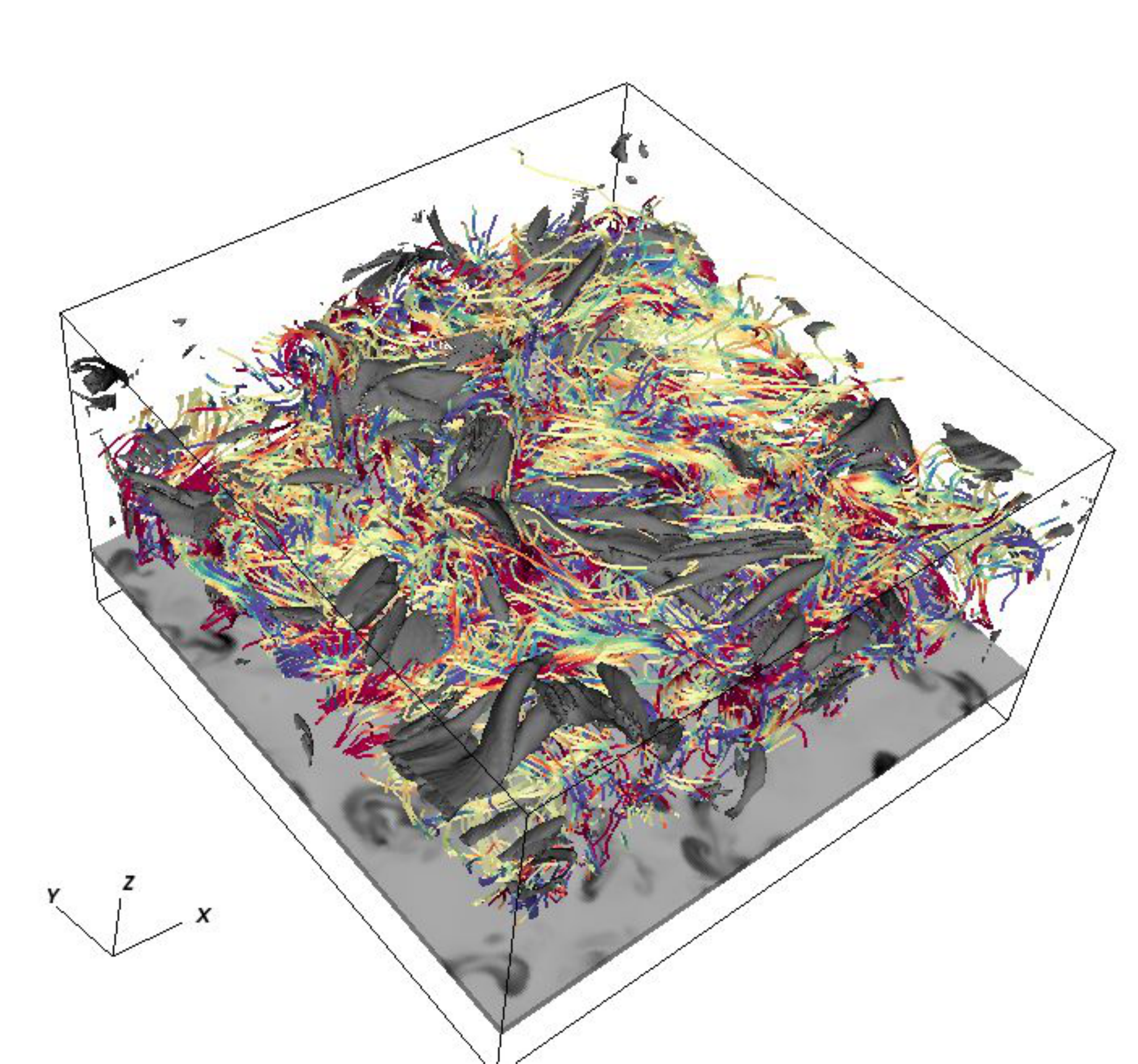}
\end{center}
\caption{Three-dimensional view of a simulation snapshot of the D-AD run. The grey-scale image at the bottom shows the temperature at $-0.5$ Mm, below the surface. Colored lines are magnetic field lines, the colors indicate the vertical magnetic field strength with blue meaning negative polarity, red for positive polarity and yellow for horizontal field. The grey contours at the upper part of the domain follow regions with the heating term log$_{10}(\eta_AJ_{\perp}^2)=7.5$. The movie showing the temporal evolution of the run is provided in the electronic version. }\label{fig:3d-bat}
\end{figure*}
\begin{figure*}
\begin{center}
\includegraphics[width = 12cm]{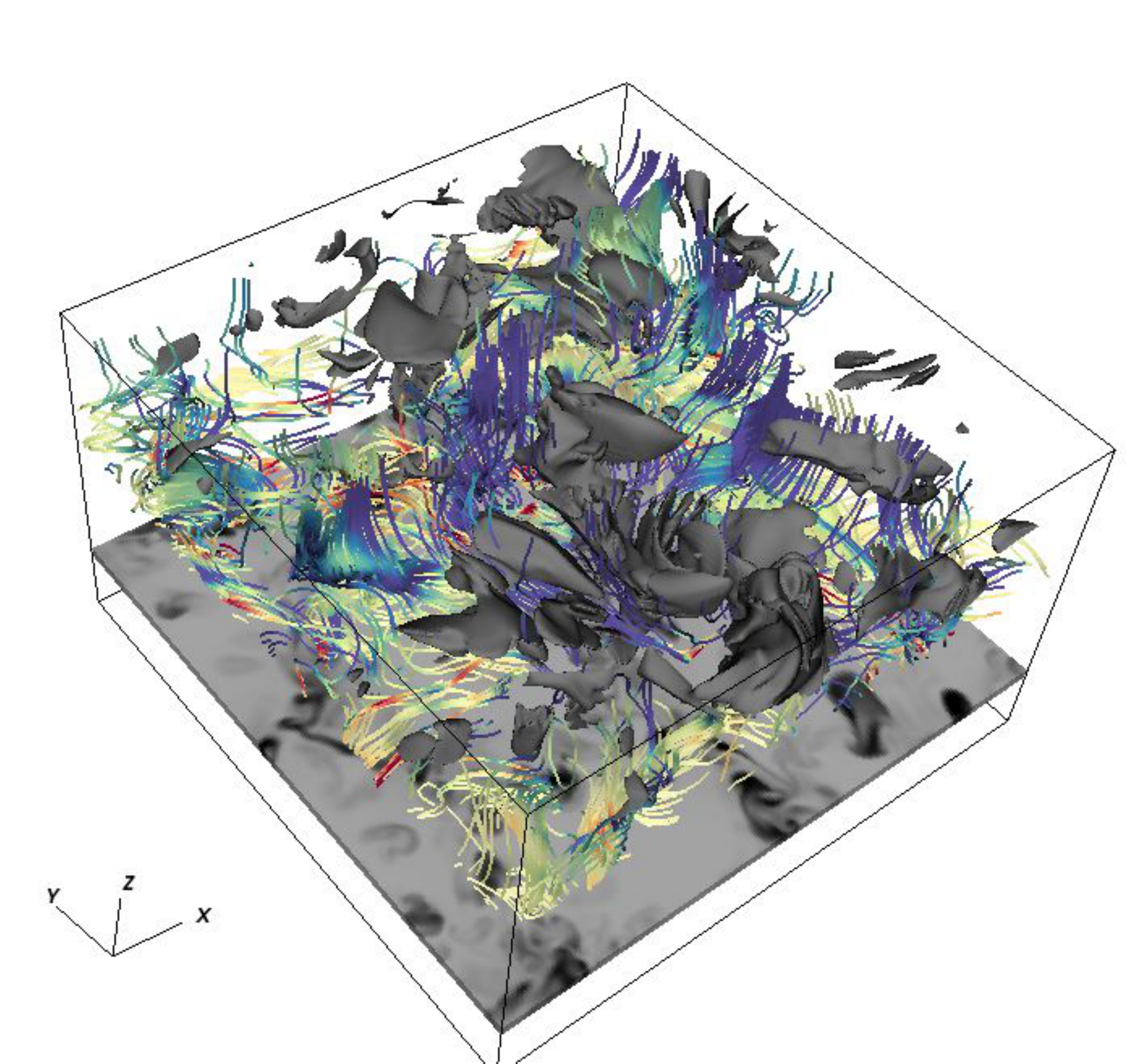}
\end{center}
\caption{Three-dimensional view of a simulation snapshot of the U-AD run. The grey-scale image at the bottom shows temperature at $-0.5$ Mm, below the surface. Colored lines are magnetic field lines, the colors indicate the vertical magnetic field strength with blue meaning negative polarity, red for positive polarity and yellow for horizontal field. The grey contours at the upper part of the domain follow regions with the heating term log$_{10}(\eta_AJ_{\perp}^2)=8$. The movie showing the temporal evolution of the run is provided in the electronic version.}\label{fig:3d-10G}
\end{figure*}

\subsection{Spatial distribution of heating}

In order to better understand why the average temperature increase is significantly lower in the U-AD simulations compared to the D-AD simulations,  Figures \ref{fig:te_heat_beta_10G} and \ref{fig:te_heat_beta_battery} show several vertical cuts of snapshots at different moments of the simulation with representative structures in both cases. Fig. \ref{fig:te_heat_beta_10G} shows the evolution of a typical vertical kG flux tube rooted in an intergranular lane and expanding with height. The structure of the flux tube can be followed by the evolution of plasma $\beta$ contours (blue lines). The plasma inside these contours has $\beta <1$ and is magnetically dominated. It can be noticed the good correlation between the locations of low plasma $\beta$ regions and large heating term regions (yellow contours) in Fig. \ref{fig:te_heat_beta_10G}. The regions where the heating term is large follow the borders of the flux tube. This is not suprising, since currents are largest at these locations. Moreover, the contours of $\eta_AJ_{\perp}^2$ correlate well with locations with low temperatures (dark red colors) at the tube borders. This correlation is especially evident in the upper layers of the simulation domain. The majority of these low-temperature regions are produced by longitudinal shocks and coincide with low density areas where the collisional coefficient $\alpha_n$ entering Eq. \ref{eq:etaa} is small leading to large values of $\eta_A$. Therefore we conclude that in the U-AD simulations with smooth and unipolar fields, the contribution of temperature effects to the heating term,  $\eta_AJ_{\perp}^2$, dominates over the contribution of the effects of currents. The heating is produced essentially in areas with low temperature  and has less impact on the mean temperature increase in the chromosphere.  

Fig. \ref{fig:te_heat_beta_battery} gives the evolution of a typical magnetic structure in the D-AD simulation. The structure of the field is significantly different in this case and no well-formed flux tube is observed. Notice that due to the generally weaker fields we do not plot the contours of log$_{10}\beta=0$, but rather  log$_{10}\beta=1$, since the former regions are rather scarce in the photosphere. Nevertheless, it can be noticed that log$_{\rm 10}\beta$ and log$_{\rm 10} \eta_A J_{\perp}^2$ contours are still well correlated, i.e. the largest heating is related to areas with small $\beta$. At the same time, there is not so good correlation between the areas with large heating term and those with low temperature, as in the U-AD case. The large heating is frequently located along the inclined canopies in the middle and upper photosphere. In the upper layers, there are still cases where low-temperature areas coincide with large $\eta_AJ_{\perp}^2$, however it must be noted that the majority of these heating events are located in the upper photospheric regions, in agreement with Fig. \ref{fig:heat}. Therefore, since the heating is mainly produced at places with no significantly lower temperatures (and densities), we conclude that it is dominated by the distribution of currents rather than by temperature effects as in the U-AD case, and this makes the overall effect on the average temperature increase more efficient. 

Yet another argument related to plasma ionization explains why adding the same amount of energy to the dense photospheric plasma results in larger effects on temperature, compared to the less dense chromospheric plasma. In the photosphere, the additional energy of the $\eta_AJ_{\perp}^2$ term is employed essentially for increasing the temperature of the plasma. However, at chromospheric heights, the energy from $\eta_AJ_{\perp}^2$ is employed for both, increasing the plasma temperature, and ionization of the plasma. Therefore, the same amount of $\eta_AJ_{\perp}^2$ in the chromosphere results in significantly less temperature increase, compared to the photosphere. Figure \ref{fig:eos} quantifies this argument using the equation of state table employed in our code. It shows cuts through the EOS table $T(e_{\rm int},\rho)$ (left panel),  and the derivative $dT/d e_{\rm int}$(right panel) for fixed values of density $\rho$, corresponding to the typical values in the photosphere and the chromosphere. Because the gradient $dT/d e_{\rm int}$ is about 5 times larger in the photosphere than in the chromosphere, the same amount of $\eta_A J^2$ leads to about 5 times larger temperature increase in the photosphere than in the chromosphere (720 vs 130 K in the case shown in the figure).

In order to further justify the above conclusions, Figures \ref{fig:mach_10G} and \ref{fig:mach_bat} present correlations between the amplitudes of the f$_{\rm long}$ and  f$_{\rm alf}$ components and the heating term at a height of 1.1 Mm (for U-AD case) and at height 0.6 Mm (for D-AD case). In order to better visualize the presence or the absence of correlation we have computed a linear fit to the maxima of distributions for a given vale of f, shown as light blue lines in the figure. This was done separately for the positive and negative values of f quantities. The behavior of the f$_{\rm fast}$ is similar to f$_{\rm long}$, and is not shown in the figures.

There is only a weak correlation between the heating term and the f$_{\rm alf}$ quantity (selecting incompressible Alfv\'en waves) in the U-AD case, Fig. \ref{fig:mach_10G}, right panel. The f$_{\rm long}$ component (selecting compressible waves with velocities along the field) does show a correlation in a way that larger heating corresponds to larger negative values of f$_{\rm long}$. The longitudinal velocity at the studied heights has a well developed non-linear behavior with shocks propagating along flux tubes. The negative values of f$_{\rm long}$ spatially coincide with rarefactions produced by shocks, implying larger instantaneous $\eta_A$, i.e. the behavior discussed in the section above. We do not expect this correlation changes even if non-instantaneous ionization is considered in the chromosphere, since the influence of density on the collisional parameter $\alpha_n$ (and $\eta_A$) is significantly more pronounced than the influence of the ionization fraction, $\xi_n$.  The parameter $\alpha_n$ changes with density on an exponential scale while the ionization fraction, $\xi_n$ in Eq. \ref{eq:etaa}, changes on a linear scale. Therefore, we conclude that the areas with a large heating term due to the AD effect coincide with rarefactions produced by longitudinal shock waves in the unipolar U-AD simulation. 

The bi-dimensional histograms for the D-AD case show a different behavior, see Fig. \ref{fig:mach_bat}. Here the f$_{\rm long}$ shows a more symmetric behavior with heating increasing with the absolute value of its amplitude. There is also a correlation for the f$_{\rm alf}$ quantity, showing an increase of the heating for positive values of  f$_{\rm alf}$ (coinciding with upflows). This behavior is just the opposite to the U-AD case. Locations with positive velocities do not correlate with the rarefaction sites, therefore the maximum heating is produced outside low density areas. This behavior confirms the visual impressions from Fig. \ref{fig:te_heat_beta_battery}.

\subsection{3D view of the structures and heating events}

Figures \ref{fig:3d-bat} and \ref{fig:3d-10G} present a three-dimensional view of the simulation domains for two selected snapshots of the D-AD and U-AD runs. The full picture of the complex interactions between the flows, magnetic field structures and the heating events can only be fully appreciated in 3D. The movies attached to the electronic version of the paper shows this complexity. 

The low-lying loops in the D-AD runs form a carpet of strongly inclined fields which determine the spatial structure of the locations with strong heating events. It can be appreciated in Fig. \ref{fig:3d-bat} that the contours with large values of $\eta_AJ_{\perp}^2$ (dark grey) follow nearly horizontal fibril-like structures formed by the magnetic fields and connecting regions with opposite polarities. These fibrils are often located above areas with the strongest fields. The movies show that their typical lifetimes are of the order of minutes. No such ``fibrils'' are observed in the upper layers of the domain since the field lines essentially connect below and do not provide sufficient currents. This is also in agreement with what was previously concluded from Fig. \ref{fig:heat}. All in all, the structuring of the heating events in horizontal fibrils resembles the observed structure of the quiet solar chromosphere.

The three-dimensional evolution in both cases clearly shows that the heating due to the ambipolar diffusion effect is intimately related to the structures formed by the magnetic fields in the Sun. The presence of structuring down to small scales, together with the action of the ambipolar diffusion, causes a larger average effect on the temperature than with the presence of strong unipolar magnetic field concentrations. These initial conclusions will need further exploration in a more detailed study.

\section{Discussion and conclusions}

This work presents the first fully three-dimensional simulations of solar magneto-convection up to chromospheric heights including one of the most important non-ideal effects due to the presence of neutrals in the solar plasma, the ambipolar diffusion. The comparison of models obtained with exactly the same numerical setup but including/excluding the ambipolar diffusion, reveals that the latter causes appreciable effects on the dynamic and thermal structure of the chromosphere in several ways. The following paragraphs summarize our main results.

Our results show that waves, excited by convective motions in the simulations, and propagating to chromospheric heights, are significantly affected by the ambipolar diffusion. We observe height and frequency dependent variations of the wave amplitudes. The variations of the amplitudes also depend on the velocity projection related to the magnetic field vector, i.e., on the wave mode. In general, more complex and structured fields of the battery-generated dynamo D-AD run cause larger effects on the wave amplitudes. 

We observe an appreciable absorption of the magnetic Poynting flux in the simulations where the ambipolar diffusion is switched on. This absorption reaches 90\% in the D-AD simulations, and it is lower, reaching up to 30\%, in the U-AD simulations. The absorption shows a weak frequency dependence and is appreciated at heights above 0.5 Mm where the heating term, $\eta_AJ_{\perp}^2$, is the largest.

The Poynting flux absorption profiles and the wave amplitude depression profiles generally do not coincide. While the amplitude maps give the amount of the available kinetic energy at a given height and frequency, the Poynting flux maps give information about the electromagnetic energy propagation, and are fundamentally different physical quantities. There can be a different explanation for the wave amplitude variations between the AD and noAD runs, such as the change of the average structure of the magnetic field caused by the ambipolar diffusion or its influence on the mode transformation process. However, the Poynting flux decrease in the AD runs compared to the noAD runs clearly tells us that the magnetic energy of the waves has been converted into heat.

The average effect of the ambipolar diffusion on the temperature structure is more pronounced in the D-AD run compared to the U-AD run. Our analysis reveals that this is caused by the existence of fields that are highly structured on small scales in the D-AD runs producing a strong ambipolar heating already over most of the photosphere, where the plasma density is high. In the U-AD runs the strong heating only exists at chromospheric heights where the plasma density is low. At chromospheric heights, the additional energy from $\eta_AJ_{\perp}^2$ term is employed not only for plasma temperature increase but also for plasma ionization, therefore its impact into the mean temperature structure is significantly less pronounced, compared to the case when the same amount of energy is added in the photosphere.

In the U-AD runs, the locations of strong heating are spatially and temporally correlated with rarefactions caused by longitudinal shock waves. This causes the correlation between the amplitude of the downward longitudinal velocity and the amplitude of the heating term, $\eta_AJ_{\perp}$. In the D-AD runs, the amplitude of the heating term is correlated with the amplitudes of the f$_{\rm alf}$ quantity (selecting incompressible waves). Therefore, the heating is produced mostly by variations of currents,  $J$ rather than $\eta_A$, causing a larger average effect on the temperature increase. The locations with strong heating are very well correlated with low plasma $\beta$ locations in the U-AD runs. This correlation is not so good for the D-AD runs with generally weaker fields. 

The three-dimensional representation of the heating events reveals that they form fibril-like structures in the D-AD simulation, which has a carpet of horizontal fields in the upper photosphere and low chromosphere. The fibrils connect regions with fields of opposite polarities and the lifetime of such fibrils is very short, of the order of one minute. Unlike that, the heating events form sheet-like structures in the U-AD simulation, following the rarefactions attached to chromospheric shocks. Their lifetimes are usually larger, of the order of several minutes. 

Some of our conclusions above reveal similar features and reinforce the conclusions from previous studies of the effects of the ambipolar diffusion on chromospheric waves, current dissipation and heating, mentioned in the introduction \citep{Goodman2000, Goodman2011, Khomenko+Collados2012, Khomenko+Collados2012b, MartinezSykora+etal2012, Shelyag+etal2016,  MartinezSykora2016,  MartinezSykora2017, Przybylski+etal2017}. In general, we observe the conversion of the magnetic energy into heat in terms of Poynting flux absorption and we confirm its effect on the average temperature structure of the chromosphere. Despite the simplicity of our chromospheric radiative transfer modeling does not allow us to make quantitative conclusions about the temperature increase, we can put a lower bound on the energization of the chromosphere due to this effect. Our D-AD runs with battery-excited dynamo fields give the basal level of magnetization in the Sun. The ambipolar heating of those areas gives us a low bound to the heating in the quietest regions of the Sun. The fact that the ambipolar diffusion shows appreciable effects on the temperature structure of the atmosphere containing only this kind of weakest fields is encouraging. It confirms the potential significance of this effect for the solution of the problem of heating of the solar chromosphere. 

The heating efficiency is proportional to the ambipolar diffusion coefficient, $\eta_A$ and to the current density perpendicular to the magnetic field, $J_{\perp}$. The former, $\eta_A$, is a function of the ionization fraction, collisional frequency and magnetic field strength. The dependence on the collision frequency is the strongest since it is proportional to density and changes exponentially with height in the solar atmosphere. Therefore, prior to analyzing the simulations presented here, we expected that the heating due to ambipolar diffusion effects would be larger in the model with stronger field structures, i.e. in the U-AD runs. Nevertheless, our analysis unexpectedly revealed the opposite. We obtain here that the efficiency of the heating is more linked to the presence of currents structured at small scales than to the variations of the neutrals fraction and density (collisional frequency). Since the ambipolar diffusion is acting at small scales \citep[see][]{Khomenko+etal2014}, the efficiency of the heating increases naturally when such currents are present, as happens in the D-AD run. In other words, the effects of small scale currents are more important than the effects of the strength of the field itself or than the effects of the change of the thermodynamic structure. This means that, even if non-equilibrium ionization and more complex radiative transfer effects are included in further modeling, we do not expect that the main conclusions of our study will change. 

Our simulations also reveal that the variation of the wave amplitude due to the ambipolar diffusion is a sensible function of their frequency and wave mode. This result indirectly confirms the conclusions by \citet{Przybylski+etal2017}, obtained in a simpler controlled experiment. Overall our simulations reveal several effects that will need separated dedicated studies using less complicated and idealized modeling, as for example the effect of ambipolar diffusion on wave transformation. The effect on the average structure of the field in the D-AD runs needs a separate dedicated study as well. 

It is also interesting to note that heating locations form fibril-like structures in the D-AD run, resembling the structuring of the quiet chromosphere. The relation between the direction of chromospheric fibrils and the magnetic field was a subject of recent studies by \citet{delaCruz2011}  who revealed that they are not always aligned. \citet{MartinezSykora2016}  confirmed such misalignment from 2D simulations including the ambipolar diffusion effects. Here we observe a similar picture. While, in the generally weaker areas with low-lying loops, fibrils tend to be aligned with the fields, this is not the case where strong unipolar tubes are present. There, the heating locations follow the shock fronts and are in general not parallel to the fields. The relation between these elongated structures seen as heating events and the actual fibrils will also need a further study.

{\bf Acknowledgements.} This work was supported by the Spanish Ministry of Science through the projects AYA2014-55078-P and AYA2014-60476-P, and by the European Research Council in the frame of the FP7 Specific Program IDEAS through the Starting Grant ERC-2011-StG 277829-SPIA. We acknowledge PRACE for awarding us access to resource MareNostrum based in Barcelona/Spain. EK acknowledges the support of the School of Mathematical Sciences at Monash University through the award of a Gordon Preston Sabbatical Fellowship.


\providecommand{\noopsort}[1]{}\providecommand{\singleletter}[1]{#1}%

\end{document}